\documentclass[12pt]{article}
\usepackage{amssymb}
\renewcommand{\theequation}{\arabic{section}.\arabic{equation}}

\begin{document}



\def\a{\alpha}
\def\b{\beta}
\def\d{\delta}
\def\e{\epsilon}
\def\g{\gamma}
\def\h{\mathfrak{h}}
\def\k{\kappa}
\def\l{\lambda}
\def\o{\omega}
\def\p{\wp}
\def\r{\rho}
\def\t{\tau}
\def\s{\sigma}
\def\z{\zeta}
\def\x{\xi}
 \def\A{{\cal{A}}}
 \def\B{{\cal{B}}}
 \def\C{{\cal{C}}}
 \def\D{{\cal{D}}}
\def\G{\Gamma}
\def\K{{\cal{K}}}
\def\O{\Omega}
\def\R{\bar{R}}
\def\T{{\cal{T}}}
\def\L{\Lambda}
\def\f{E_{\tau,\eta}(sl_2)}
\def\E{E_{\tau,\eta}(sl_n)}
\def\Zb{\mathbb{Z}}
\def\Cb{\mathbb{C}}

\def\R{\overline{R}}

\def\beq{\begin{equation}}
\def\eeq{\end{equation}}
\def\bea{\begin{eqnarray}}
\def\eea{\end{eqnarray}}
\def\ba{\begin{array}}
\def\ea{\end{array}}
\def\no{\nonumber}
\def\le{\langle}
\def\re{\rangle}
\def\lt{\left}
\def\rt{\right}

\newtheorem{Theorem}{Theorem}
\newtheorem{Definition}{Definition}
\newtheorem{Proposition}{Proposition}
\newtheorem{Lemma}{Lemma}
\newtheorem{Corollary}{Corollary}
\newcommand{\proof}[1]{{\bf Proof. }
        #1\begin{flushright}$\Box$\end{flushright}}

\baselineskip=20pt

\newfont{\elevenmib}{cmmib10 scaled\magstep1}
\newcommand{\preprint}{
   \begin{flushleft}
   \end{flushleft}\vspace{-1.3cm}
   \begin{flushright}\normalsize
   \end{flushright}}
\newcommand{\Title}[1]{{\baselineskip=26pt
   \begin{center} \Large \bf #1 \\ \ \\ \end{center}}}
\newcommand{\Author}{\begin{center}
   \large \bf
Wen-Li Yang${}^{a}$
 ~and~ Yao-Zhong Zhang ${}^b$
 \end{center}}
\newcommand{\Address}{\begin{center}

     ${}^a$ Institute of Modern Physics, Northwest University,
     Xian 710069, P.R. China\\
     ${}^b$ The University of Queensland, School of Mathematics and Physics,  Brisbane, QLD 4072,
     Australia\\
    E-mail: wlyang@nwu.edu.cn,\,\, yzz@maths.uq.edu.au
   \end{center}}
\newcommand{\Accepted}[1]{\begin{center}
   {\large \sf #1}\\ \vspace{1mm}{\small \sf Accepted for Publication}
   \end{center}}

\preprint
\thispagestyle{empty}
\bigskip\bigskip\bigskip

\Title{Drinfeld twists of the open XXZ chain with non-diagonal
boundary terms } \Author

\Address
\vspace{1cm}

\begin{abstract}
The Drinfeld twists or factorizing F-matrices for the open XXZ
spin chain with non-diagonal boundary terms are constructed. It is
shown that in the F-basis the two sets of pseudo-particle creation
operators simultaneously take completely symmetric and
polarization free form. The explicit and completely symmetric
expressions of the two sets of Bethe states of the model are
obtained.

\vspace{1truecm} \noindent {\it PACS:} 03.65.Fd; 04.20.Jb;
05.30.-d; 75.10.Jm

\end{abstract}
\newpage
\section{Introduction}
\label{intro} \setcounter{equation}{0}

The algebraic Bethe ansatz \cite{Kor93} has been proven to provide
a powerful tool of solving eigenvalue problems of quantum
integrable systems such as quantum spin chains. In this framework,
the pseudo-particle creation and annihilation operators are
constructed by off-diagonal matrix elements of the so-called
monodromy matrix. The Bethe states (eigenstates) of transfer
matrix are obtained by applying creation operators to the
reference state (or pseudo-vacuum state). However, the apparently
simple action of creation operators is plagued with non-local
effects arising from polarization clouds or compensating exchange
terms on the level of local operators. This make the explicit
construction of the Bethe states challenging.

In \cite{Mai00}, Maillet and Sanchez de Santos  showed how
monodromy matrices of the inhomogeneous XXX and XXZ spin chains
with periodic boundary conditions (or the closed XXX and XXZ
chains) can be simplified by using the Drinfeld twists or
factorizing F-matrices \cite{Dri83}. This leads to the so-called
F-basis provided by the F-matrix for the analysis of these models.
In the F-basis, the pseudo-particle creation and annihilation
operators  of the models  take completely symmetric forms and
contain no compensating exchange terms on the level of local
operators (i.e. polarization free). As a result, the Bethe states
of the models are simplified dramatically and can be written down
explicitly \cite{Kit99}. Similar results have been obtained for
other models with periodic boundary conditions
\cite{Ter99,Alb00,Alb00-1,Alb01,Zha06,Yan06-1}.

Recently, it was  shown \cite{Wan02,Kit07} that the F-matrices of
the closed XXX and XXZ chains also  make the pseudo-particle
creation operators of the open XXX and XXZ chains with diagonal
boundary terms polarization free. This is mainly due to the fact
that the closed chain and the corresponding open chain with
diagonal boundary terms share the same reference state
\cite{Skl88}. However, the story for the open XXZ chain with
non-diagonal boundary terms is quite different
\cite{Nep04,Cao03,Yan04,Gal05,Gie05,Gie05-1,Yan04-1,Baj06,Yan05,Doi06,Mur06,Bas07,Gal08,Mur09}.
Firstly, the reference state (all spin up state) of the closed
chain is no longer a reference state of the open chain with
non-diagonal boundary terms \cite{Cao03,Yan04,Yan04-1}. Secondly,
two reference states (and thus two sets of Bethe states) are
needed \cite{Yan07} for the open XXZ chain with non-diagonal
boundary terms in order to obtain its complete spectrum
\cite{Nep03,Yan06}. As a consequence, the F-matrix found in
\cite{Mai00} is no longer the {\it desirable} F-matrix for the
open XXZ chain with non-diagonal boundary terms.

In this paper, we obtain the Drinfeld twists or factorizing
F-matrices for the open XXZ chain with integrable boundary
conditions given  by the non-diagonal K-matrices
(\ref{K-matrix-2-1}) and (\ref{K-matrix-6}). We find that in the
F-basis the two sets of pseudo-particle creation operators (acting
on the two reference states) of the boundary model simultaneously
take completely symmetric and polarization free forms. These
enable us to derive the explicit and completely symmetric
expressions of the two sets of Bethe states of the model.

The paper is organized as follows.  In section 2, we briefly
describe the open XXZ chain with non-diagonal boundary terms and
introduce the pseudo-particle creation operators and the two sets
of Bethe states of the model. In section 3, we introduce a new
pseudo-particle creation operator. This operator is used to
generate a new set of Bethe states. Section 4 introduces the
one-row and double row monodromy matrices in the face picture for
the open XXZ chain. In section 5, we construct the F-matrix of the
open XXZ chain in the face picture and obtain  the completely
symmetric and polarization free representations of the
pseudo-particle creation operators and the complete symmetric
expressions of the two sets of Bethe states in the F-basis. In
section 6, we summarize our results and give some discussions.
Some details about the new set of Bethe states are given in
Appendix A.


\section{ The inhomogeneous spin-$\frac{1}{2}$ XXZ open chain}
\label{XXZ} \setcounter{equation}{0}

Throughout, $V$ denotes a two-dimensional linear space. The
spin-$\frac{1}{2}$ XXZ chain can be constructed from the
well-known six-vertex model R-matrix $\R(u)\in {\rm End}(V\otimes
V)$ \cite{Kor93} given by \bea
\R(u)=\lt(\begin{array}{llll}1&&&\\&b(u)&c(u)&\\
&c(u)&b(u)&\\&&&1\end{array}\rt).\label{r-matrix}\eea The
coefficient functions read: $b(u)=\frac{\sin u}{\sin(u+\eta)}$,
$c(u)=\frac{\sin\eta}{\sin(u+\eta)}$. Here we assume  $\eta$ is a
generic complex number. The R-matrix satisfies the quantum
Yang-Baxter equation (QYBE), \bea
R_{1,2}(u_1-u_2)R_{1,3}(u_1-u_3)R_{2,3}(u_2-u_3)
=R_{2,3}(u_2-u_3)R_{1,3}(u_1-u_3)R_{1,2}(u_1-u_2),\label{QYB}\eea
and the unitarity, crossing-unitarity and quasi-classical
properties \cite{Yan04-1}. We adopt the standard notations: for
any matrix $A\in {\rm End}(V)$ , $A_j$ (or $A^j$) is an embedding
operator in the tensor space $V\otimes V\otimes\cdots$, which acts
as $A$ on the $j$-th space and as identity on the other factor
spaces; $R_{i,j}(u)$ is an embedding operator of R-matrix in the
tensor space, which acts as identity on the factor spaces except
for the $i$-th and $j$-th ones.

One introduces the ``row-to-row"  (or one-row ) monodromy matrix
$T(u)$, which is an $2\times 2$ matrix with elements being
operators acting on $V^{\otimes N}$, where $N=2M$ ($M$ being a
positive integer),\bea
T_0(u)=\R_{0,N}(u-z_N)\R_{0,N-1}(u-z_{N-1})\cdots
\R_{0,1}(u-z_1).\label{Mon-V}\eea Here $\{z_j|j=1,\cdots,N\}$ are
arbitrary free complex parameters which are usually called
inhomogeneous parameters.

Integrable open chain can be constructed as follows \cite{Skl88}.
Let us introduce a pair of K-matrices $K^-(u)$ and $K^+(u)$. The
former satisfies the reflection equation (RE)
 \bea &&\R_{1,2}(u_1-u_2)K^-_1(u_1)\R_{2,1}(u_1+u_2)K^-_2(u_2)\no\\
 &&~~~~~~=
K^-_2(u_2)\R_{1,2}(u_1+u_2)K^-_1(u_1)\R_{2,1}(u_1-u_2),\label{RE-V}\eea
and the latter  satisfies the dual RE \bea
&&\R_{1,2}(u_2-u_1)K^+_1(u_1)\R_{2,1}(-u_1-u_2-2\eta)K^+_2(u_2)\no\\
&&~~~~~~=
K^+_2(u_2)\R_{1,2}(-u_1-u_2-2\eta)K^+_1(u_1)\R_{2,1}(u_2-u_1).
\label{DRE-V}\eea For open spin-chains, instead of the standard
``row-to-row" monodromy matrix $T(u)$ (\ref{Mon-V}), one needs to
consider  the
 ``double-row" monodromy matrix $\mathbb{T}(u)$
\bea
  \mathbb{T}(u)=T(u)K^-(u)\hat{T}(u),\quad \hat{T}(u)=T^{-1}(-u).
  \label{Mon-V-0}
\eea Then the double-row transfer matrix of the XXZ chain with
open boundary (or the open XXZ chain) is given by \bea
\t(u)=tr(K^+(u)\mathbb{T}(u)).\label{trans}\eea The QYBE and
(dual) REs lead to that the transfer matrices with different
spectral parameters commute with each other \cite{Skl88}:
$[\t(u),\t(v)]=0$. This ensures the integrability of the open XXZ
chain.

In this paper, we will consider the K-matrix $K^{-}(u)$ which is a
generic solution to the RE (\ref{RE-V}) associated the six-vertex
model R-matrix  \cite{Veg93,Gho94}
\bea K^-(u)=\lt(\begin{array}{ll}k_1^1(u)&k^1_2(u)\\
k^2_1(u)&k^2_2(u)\end{array}\rt)\equiv K(u).\label{K-matrix}\eea
The coefficient functions are \bea && k^1_1(u)=
\frac{\cos(\l_1-\l_2) -\cos(\l_1+\l_2+2\xi)e^{-2iu}}
{2\sin(\l_1+\xi+u)
\sin(\l_2+\xi+u)},\no\\
&&k^1_2(u)=\frac{-i\sin(2u)e^{-i(\l_1+\l_2)} e^{-iu}}
{2\sin(\l_1+\xi+u) \sin(\l_2+\xi+u)},\no\\
&&k^2_1(u)=\frac{i\sin(2u)e^{i(\l_1+\l_2)} e^{-iu}}
{2\sin(\l_1+\xi+u) \sin(\l_2+\xi+u)}, \no\\
&& k^2_2(u)=\frac{\cos(\l_1-\l_2)e^{-2iu}- \cos(\l_1+\l_2+2\xi)}
{2\sin(\l_1+\xi+u)\sin(\l_2+\xi+u)}.\label{K-matrix-2-1} \eea At
the same time, we introduce  the corresponding {\it dual\/}
K-matrix $K^+(u)$ which is a generic solution to the dual
reflection equation (\ref{DRE-V}) with a particular choice of the
free boundary parameters:
\bea K^+(u)=\lt(\begin{array}{ll}{k^+}_1^1(u)&{k^+}^1_2(u)\\
{k^+}^2_1(u)&{k^+}^2_2(u)\end{array}\rt)\label{DK-matrix}\eea with
the matrix elements \bea && {k^+}^1_1(u)=
\frac{\cos(\l_1-\l_2)e^{-i\eta}
-\cos(\l_1+\l_2+2\bar{\xi})e^{2iu+i\eta}}
{2\sin(\l_1+\bar{\xi}-u-\eta)
\sin(\l_2+\bar{\xi}-u-\eta)},\no\\
&&{k^+}^1_2(u)=\frac{i\sin(2u+2\eta)e^{-i(\l_1+\l_2)}
e^{iu-i\eta}} {2\sin(\l_1+\bar{\xi}-u-\eta)
\sin(\l_2+\bar{\xi}-u-\eta)},
\no\\
&&{k^+}^2_1(u)=\frac{-i\sin(2u+2\eta)e^{i(\l_1+\l_2)}
e^{iu+i\eta}}
{2\sin(\l_1+\bar{\xi}-u-\eta) \sin(\l_2+\bar{\xi}-u-\eta)}, \no\\
&& {k^+}^2_2(u)=\frac{\cos(\l_1-\l_2)e^{2iu+i\eta}-
\cos(\l_1+\l_2+2\bar{\xi})e^{-i\eta}}
{2\sin(\l_1+\bar{\xi}-u-\eta)\sin(\l_2+\bar{\xi}-u-\eta)}.\label{K-matrix-6}
\eea The K-matrices depend on four free boundary parameters
$\{\l_1,\,\l_2,\,\xi,\,\bar{\xi}\}$ \footnote{To our knowledge,
this is the only case for which the complete eigenstates of the
open XXZ chain with non-diagonal boundary terms can be constructed
through the algebraic Bethe ansatz method \cite{Yan07}.}. It is
very convenient to introduce a vector $\l\in V$ associated with
the boundary parameters $\{\l_i\}$, \bea
 \l=\sum_{k=1}^2\l_k\e_k, \label{boundary-vector}
\eea  where $\{\e_i,\,i=1,2\}$ form the orthonormal basis of $V$
such that $\langle \e_i,\e_j\rangle=\d_{ij}$.


\subsection{Vertex-face correspondence}

Let us briefly review the face-type R-matrix associated with the
six-vertex model.

Set \bea \hat{\imath}=\e_i-\overline{\e},~~\overline{\e}=
\frac{1}{2}\sum_{k=1}^{2}\e_k, \quad i=1,2,\qquad {\rm then}\,
\sum_{i=1}^2\hat{\imath}=0. \label{fundmental-vector} \eea Let
$\h$ be the Cartan subalgebra of $A_{1}$ and $\h^{*}$ be its dual.
A finite dimensional diagonalizable  $\h$-module is a complex
finite dimensional vector space $W$ with a weight decomposition
$W=\oplus_{\mu\in \h^*}W[\mu]$, so that $\h$ acts on $W[\mu]$ by
$x\,v=\mu(x)\,v$, $(x\in \h,\,v\in\,W[\mu])$. For example, the
non-zero weight spaces of the fundamental representation
$V_{\L_1}=\Cb^2=V$ are
\bea
 W[\hat{\imath}]=\Cb \e_i,~i=1,2.\label{Weight}
\eea

For a generic $m\in V$, define \bea m_i=\langle m,\e_i\rangle,
~~m_{ij}=m_i-m_j=\langle m,\e_i-\e_j\rangle,~~i,j=1,2.
\label{Def1}\eea Let $R(u,m)\in {\rm End}(V\otimes V)$ be the
R-matrix of the six-vertex SOS model, which is trigonometric limit
of the eight-vertex SOS model \cite{Bax82} given by
\bea
R(u;m)\hspace{-0.1cm}=\hspace{-0.1cm}
\sum_{i=1}^{2}R(u;m)^{ii}_{ii}E_{ii}\hspace{-0.1cm}\otimes\hspace{-0.1cm}
E_{ii}\hspace{-0.1cm}+\hspace{-0.1cm}\sum_{i\ne
j}^2\lt\{R(u;m)^{ij}_{ij}E_{ii}\hspace{-0.1cm}\otimes\hspace{-0.1cm}
E_{jj}\hspace{-0.1cm}+\hspace{-0.1cm}
R(u;m)^{ji}_{ij}E_{ji}\hspace{-0.1cm}\otimes\hspace{-0.1cm}
E_{ij}\rt\}, \label{R-matrix} \eea where $E_{ij}$ is the matrix
with elements $(E_{ij})^l_k=\d_{jk}\d_{il}$. The coefficient
functions are \bea
 &&R(u;m)^{ii}_{ii}=1,~~
   R(u;m)^{ij}_{ij}=\frac{\sin u\sin(m_{ij}-\eta)}
   {\sin(u+\eta)\sin(m_{ij})},~~i\neq j,\label{Elements1}\\
 && R(u;m)^{ji}_{ij}=\frac{\sin\eta\sin(u+m_{ij})}
    {\sin(u+\eta)\sin(m_{ij})},~~i\neq j,\label{Elements2}
\eea  and $m_{ij}$ is defined in (\ref{Def1}). The R-matrix
satisfies the dynamical (modified) quantum Yang-Baxter equation
(or the star-triangle relation) \cite{Bax82}
\begin{eqnarray}
&&R_{1,2}(u_1-u_2;m-\eta h^{(3)})R_{1,3}(u_1-u_3;m)
R_{2,3}(u_2-u_3;m-\eta h^{(1)})\no\\
&&\qquad =R_{2,3}(u_2-u_3;m)R_{1,3}(u_1-u_3;m-\eta
h^{(2)})R_{1,2}(u_1-u_2;m).\label{MYBE}
\end{eqnarray}
Here we have adopted
\bea R_{1,2}(u,m-\eta h^{(3)})\,v_1\otimes
v_2 \otimes v_3=\lt(R(u,m-\eta\mu)\otimes {\rm id }\rt)v_1\otimes
v_2 \otimes v_3,\quad {\rm if}\, v_3\in W[\mu]. \label{Action}
\eea Moreover, one may check that the R-matrix satisfies  weight
conservation condition, \bea
  \lt[h^{(1)}+h^{(2)},\,R_{1,2}(u;m)\rt]=0,\label{Conservation}
\eea unitary condition, \bea
 R_{1,2}(u;m)\,R_{2,1}(-u;m)={\rm id}\otimes {\rm
 id},\label{Unitary}
\eea and crossing relation \bea
 R(u;m)^{kl}_{ij}=\varepsilon_{l}\,\varepsilon_{j}
   \frac{\sin(u)\sin((m-\eta\hat{\imath})_{21})}
   {\sin(u+\eta)\sin(m_{21})}R(-u-\eta;m-\eta\hat{\imath})
   ^{\bar{j}\,k}_{\bar{l}\,i},\label{Crossing}
\eea where
\bea \varepsilon_{1}=1,\,\varepsilon_{2}=-1,\quad {\rm
and}\,\, \bar{1}=2,\,\bar{2}=1.\label{Parity} \eea

Define the following functions: $\theta^{(1)}(u)=e^{-iu}$, $
\theta^{(2)}(u)=1$. Let us introduce two intertwiners which are
$2$-component  column vectors $\phi_{m,m-\eta\hat{\jmath}}(u)$
labelled by $\hat{1},\,\hat{2}$. The $k$-th element of
$\phi_{m,m-\eta\hat{\jmath}}(u)$ is given by \bea
\phi^{(k)}_{m,m-\eta\hat{\jmath}}(u)=\theta^{(k)}(u+2m_j).\label{Intvect}\eea
Explicitly, \bea \phi_{m,m-\eta\hat{1}}(u)=
\lt(\begin{array}{c}e^{-i(u+2m_1)}\\1\end{array}\rt),\qquad
\phi_{m,m-\eta\hat{2}}(u)=
\lt(\begin{array}{c}e^{-i(u+2m_2)}\\1\end{array}\rt).\eea
Obviously, the two intertwiner vectors
$\phi_{m,m-\eta\hat{\imath}}(u)$ are linearly {\it independent}
for a generic $m\in V$.

 Using the intertwiner vectors, one can derive the following face-vertex
correspondence relation \cite{Cao03}\bea &&\R_{1,2}(u_1-u_2)
\phi^1_{m,m-\eta\hat{\imath}}(u_1)
\phi^2_{m-\eta\hat{\imath},m-\eta(\hat{\imath}+\hat{\jmath})}(u_2)
\no\\&&~~~~~~= \sum_{k,l}R(u_1-u_2;m)^{kl}_{ij}
\phi^1_{m-\eta\hat{l},m-\eta(\hat{l}+\hat{k})}(u_1)
\phi^2_{m,m-\eta\hat{l}}(u_2). \label{Face-vertex} \eea  Then the
QYBE (\ref{QYB}) of the vertex-type R-matrix $\R(u)$ is equivalent
to the dynamical Yang-Baxter equation (\ref{MYBE}) of the SOS
R-matrix $R(u,m)$. For a generic $m$, we can introduce other types
of intertwiners $\bar{\phi},~\tilde{\phi}$ which  are both row
vectors and satisfy the following conditions, \bea
  &&\bar{\phi}_{m,m-\eta\hat{\mu}}(u)
     \,\phi_{m,m-\eta\hat{\nu}}(u)=\d_{\mu\nu},\quad
     \tilde{\phi}_{m+\eta\hat{\mu},m}(u)
     \,\phi_{m+\eta\hat{\nu},m}(u)=\d_{\mu\nu},\label{Int2}\eea
{}from which one  can derive the relations,
\begin{eqnarray}
&&\sum_{\mu=1}^2\phi_{m,m-\eta\hat{\mu}}(u)\,
 \bar{\phi}_{m,m-\eta\hat{\mu}}(u)={\rm id},\label{Int3}\\
&&\sum_{\mu=1}^2\phi_{m+\eta\hat{\mu},m}(u)\,
 \tilde{\phi}_{m+\eta\hat{\mu},m}(u)={\rm id}.\label{Int4}
\end{eqnarray}
With the help of (\ref{Face-vertex})-(\ref{Int4}), we obtain,
\begin{eqnarray}
 &&\tilde{\phi}^1_{m+\eta\hat{k},m}(u_1)\,\R_{1,2}(u_1-u_2)
 \phi^2_{m+\eta\hat{\jmath},m}(u_2)\no\\
 &&\qquad\quad= \sum_{i,l}R(u_1-u_2;m)^{kl}_{ij}\,
 \tilde{\phi}^1_{m+\eta(\hat{\imath}+\hat{\jmath}),m+\eta\hat{\jmath}}(u_1)
 \phi^2_{m+\eta(\hat{k}+\hat{l}),m+\eta\hat{k}}(u_2),\label{Face-vertex1}\\
 &&\tilde{\phi}^1_{m+\eta\hat{k},m}(u_1)
 \tilde{\phi}^2_{m+\eta(\hat{k}+\hat{l}),m+\eta\hat{k}}(u_2)\,
 \R_{1,2}(u_1-u_2)\no\\
 &&\qquad\quad= \sum_{i,j}R(u_1-u_2;m)^{kl}_{ij}\,
 \tilde{\phi}^1_{m+\eta(\hat{\imath}+\hat{\jmath}),m+\eta\hat{\jmath}}(u_1)
 \tilde{\phi}^2_{m+\eta\hat{\jmath},m}(u_2),\label{Face-vertex2}\\
 &&\bar{\phi}^2_{m,m-\eta\hat{l}}(u_2)\,\R_{1,2}(u_1-u_2)
   \phi^1_{m,m-\eta\hat{\imath}}(u_1)\no\\
 &&\qquad\quad= \sum_{k,j}R(u_1-u_2;m)^{kl}_{ij}\,
 \phi^1_{m-\eta\hat{l},m-\eta(\hat{k}+\hat{l})}(u_1)
 \bar{\phi}^2_{m-\eta\hat{\imath},m-\eta(\hat{\imath}+\hat{\jmath})}(u_2),\label{Face-vertex3}\\
 &&\bar{\phi}^1_{m-\eta\hat{l},m-\eta(\hat{k}+\hat{l})}(u_1)
 \bar{\phi}^2_{m,m-\eta\hat{l}}(u_2)\,\R_{12}(u_1-u_2)\no\\
 &&\qquad\quad= \sum_{i,j}R(u_1-u_2;m)^{kl}_{ij}\,
 \bar{\phi}^1_{m,m-\eta\hat{\imath}}(u_1)
 \bar{\phi}^2_{m-\eta\hat{\imath},m-\eta(\hat{\imath}
 +\hat{\jmath})}(u_2).\label{Face-vertex4}
\end{eqnarray}

One may verify that the K-matrices $K^{\pm}(u)$ given by
(\ref{K-matrix}) and (\ref{DK-matrix}) can be expressed in terms
of the intertwiners and {\it diagonal\/} matrices $\K(\l|u)$ and
$\tilde{\K}(\l|u)$ as follows \bea &&K^-(u)^s_t=
\sum_{i,j}\phi^{(s)}_{\l-\eta(\hat{\imath}-\hat{\jmath}),
~\l-\eta\hat{\imath}}(u)
\K(\l|u)^j_i\bar{\phi}^{(t)}_{\l,~\l-\eta\hat{\imath}}(-u),\label{K-F-1}\\
&&K^+(u)^s_t= \sum_{i,j}
\phi^{(s)}_{\l,~\l-\eta\hat{\jmath}}(-u)\tilde{\K}(\l|u)^j_i
\tilde{\phi}^{(t)}_{\l-\eta(\hat{\jmath}-\hat{\imath}),
~\l-\eta\hat{\jmath}}(u).\label{K-F-2}\eea Here the two {\it
diagonal\/} matrices $\K(\l|u)$ and $\tilde{\K}(\l|u)$ are given
by \bea
&&\K(\l|u)\equiv{\rm Diag}(k(\l|u)_1,\,k(\l|u)_2)={\rm
Diag}(\frac{\sin(\l_1+\xi-u)}{\sin(\l_1+\xi+u)},\,
\frac{\sin(\l_2+\xi-u)}{\sin(\l_2+\xi+u)}),\label{K-F-3}\\
&&\tilde{\K}(\l|u)\equiv{\rm
Diag}(\tilde{k}(\l|u)_1,\,\tilde{k}(\l|u)_2)\no\\
&&~~~~~~~~~={\rm
Diag}(\frac{\sin(\l_{12}\hspace{-0.1cm}-\hspace{-0.1cm}
\eta)\sin(\l_1\hspace{-0.1cm}+\hspace{-0.1cm}\bar{\xi}+\hspace{-0.1cm}u
\hspace{-0.1cm}+\hspace{-0.1cm}\eta)}
{\sin\l_{12}\sin(\l_1+\bar{\xi}-u-\eta)},\,
\frac{\sin(\l_{12}\hspace{-0.1cm}+\hspace{-0.1cm}
\eta)\sin(\l_2\hspace{-0.1cm}+\hspace{-0.1cm}\bar{\xi}\hspace{-0.1cm}
+\hspace{-0.1cm}u\hspace{-0.1cm}+\hspace{-0.1cm}\eta)}
{\sin\l_{12}\sin(\l_2+\bar{\xi}-u-\eta)}).\label{K-F-4} \eea
Although the vertex type K-matrices $K^{\pm}(u)$ given by
(\ref{K-matrix}) and (\ref{DK-matrix}) are generally non-diagonal,
after the face-vertex transformations (\ref{K-F-1}) and
(\ref{K-F-2}), the face type counterparts $\K(\l|u)$ and
$\tilde{\K}(\l|u)$  become {\it simultaneously\/} diagonal. This
fact enabled the authors to apply the generalized algebraic Bethe
ansatz method developed in \cite{Yan04} for SOS type integrable
models to diagonalize the transfer matrices $\t(u)$ (\ref{trans})
\cite{Yan04-1,Yan07}.

\subsection{Two sets of eigenstates}

By (\ref{K-F-2}) and (\ref{K-F-4}),  we can recast the transfer
matrix $\t(u)$ (\ref{trans}) in the form \bea
&&\t(u)=tr(K^+(u)\mathbb{T}(u))
=\sum_{\mu,\nu}\tilde{\K}(\l|u)_{\nu}^{\mu}\T^-(\l|u)^{\nu}_{\mu}=
\sum_{\mu}\tilde{k}(\l|u)_{\mu}\T^-(\l|u)^{\mu}_{\mu}.
\label{De1}\eea Here we have introduced the new double-row
monodromy matrix $\T^-(m|u)$
\bea
 \T^-(m|u)^{\nu}_{\mu}
  =\tilde{\phi}^{0}_{m-\eta(\hat{\mu}-\hat{\nu}),
  m-\eta\hat{\mu}}(u)~\mathbb{T}_0(u)\phi^{0}_{m,
  m-\eta\hat{\mu}}(-u).\label{Mon-F}
\eea This double-row monodromy matrix, in the face picture, can be
expressed in terms of the face type R-matrix $R(u;m)$
(\ref{R-matrix}) and K-matrix $\K(\l|u)$ (\ref{K-F-3}) (for the
details, see  (\ref{Expression-3}) below).

In contrast with the case of  diagonal boundary terms
\cite{Skl88}, there exist  two sets of Bethe states which
constitute the complete set of eigenstates of the transfer matrix
for the models with non-diagonal boundary terms \cite{Yan07}.
These two sets of states are given by \bea
&&|v^{(1)}_1,\cdots,v^{(1)}_M\rangle^{(1)}=
     \T^-(\l-2\eta\hat{1}|v^{(1)}_1)^1_2
   \cdots
   \T^-(\l-2M\eta\hat{1}|v^{(1)}_M)^1_2|\O^{(1)}(\l)\rangle,
   \label{Bethe-state}\\
&&|v^{(2)}_1,\cdots,v^{(2)}_M\rangle^{(II)} =
   \T^-(\l-2\eta\hat{2}|v^{(2)}_1)^2_1
   \cdots
   \T^-(\l\hspace{-0.04truecm}-\hspace{-0.04truecm}2M\eta\hat{2}|v^{(2)}_M)^2_1|\O^{(II)}(\l)\rangle,
   \label{Bethe-state-2}
\eea where the vector $\l$ is related to the boundary parameters
(\ref{boundary-vector}). The associated reference states
$|\O^{(1)}(\l)\rangle$ and $|\O^{(II)}(\l)\rangle$ are \bea
\hspace{-1.2truecm}|\O^{(1)}(\l)\rangle
&=&\phi^1_{\l,\l-\eta\hat{1}}(z_1)
\phi^{2}_{\l-\eta\hat{1},\l-2\eta\hat{1}}(z_{2})\cdots
\phi^N_{\l-(N-1)\eta\hat{1},\l-N\eta\hat{1}}(z_N)
,\label{Vac}\\
\hspace{-1.2truecm} |\O^{(II)}(\l)\rangle&=&
\phi^1_{\l,\l-\eta\hat{2}}(z_1)
\phi^{2}_{\l-\eta\hat{2},\l-2\eta\hat{2}}(z_{2})\cdots
\phi^N_{\l-(N-1)\eta\hat{2},\l-N\eta\hat{2}}(z_N).\label{Vac-2}
\eea It is remarked that   $\phi^k={\rm id}\otimes {\rm
id}\cdots\otimes \stackrel{k-th}{\phi}\otimes {\rm id}\cdots$.

If the parameters $\{v^{(1)}_k\}$ satisfy the first set of  Bethe
ansatz equations, \bea &&\hspace{-0.1cm}\frac
{\sin(\l_2+\xi+v^{(1)}_{\a})\sin(\l_2+\bar\xi-v^{(1)}_{\a})
\sin(\l_1+\bar\xi+v^{(1)}_{\a})\sin(\l_1+\xi-v^{(1)}_{\a})}
{\sin(\l_2\hspace{-0.1cm}+\hspace{-0.1cm}
\bar\xi\hspace{-0.1cm}+\hspace{-0.1cm}v^{(1)}_{\a}
\hspace{-0.1cm}+\hspace{-0.1cm}\eta)
\sin(\l_2\hspace{-0.1cm}+\hspace{-0.1cm}\xi\hspace{-0.1cm}-\hspace{-0.1cm}v^{(1)}_{\a}
\hspace{-0.1cm}-\hspace{-0.1cm}\eta)
\sin(\l_1\hspace{-0.1cm}+\hspace{-0.1cm}\xi\hspace{-0.1cm}+\hspace{-0.1cm}
v^{(1)}_{\a}\hspace{-0.1cm}+\hspace{-0.1cm}\eta)
\sin(\l_1\hspace{-0.1cm}+\hspace{-0.1cm}\bar\xi\hspace{-0.1cm}-\hspace{-0.1cm}v^{(1)}_{\a}
\hspace{-0.1cm}-\hspace{-0.1cm}\eta)}\no\\
&&~~~~~~=\prod_{k\neq
\a}^M\frac{\sin(v^{(1)}_{\a}+v^{(1)}_k+2\eta)\sin(v^{(1)}_{\a}-v^{(1)}_k+\eta)}
{\sin(v^{(1)}_{\a}+v^{(1)}_k)\sin(v^{(1)}_{\a}-v^{(1)}_k-\eta)}\no\\
&&~~~~~~~~~~\times\prod_{k=1}^{2M}\frac{\sin(v^{(1)}_{\a}+z_k)\sin(v^{(1)}_{\a}-z_k)}
{\sin(v^{(1)}_{\a}+z_k+\eta)\sin(v^{(1)}_{\a}-z_k+\eta)},~~\a=1,\cdots,M,
\label{BA-D-1}\eea the Bethe state
$|v^{(1)}_1,\cdots,v^{(1)}_M\rangle^{(1)}$ becomes the eigenstate
of the transfer matrix with eigenvalue $\L^{(1)}(u)$  given by
\cite{Yan04-1}
\bea
&&\L^{(1)}(u)=\frac{\sin(\l_2+\bar\xi-u)\sin(\l_1+\bar\xi+u)\sin(\l_1+\xi-u)\sin(2u+2\eta)}
{\sin(\l_2+\bar\xi-u-\eta)\sin(\l_1+\bar\xi-u-\eta)\sin(\l_1+\xi+u)\sin(2u+\eta)}\no\\
&&~~~~~~~~~~~~~~~~~~\times\prod_{k=1}^M\frac{\sin(u+v^{(1)}_k)\sin(u-v^{(1)}_k-\eta)}
{\sin(u+v^{(1)}_k+\eta)\sin(u-v^{(1)}_k)}\no\\
&&~~~~~~+\frac{\sin(\l_2+\bar\xi+u+\eta)\sin(\l_1+\xi+u+\eta)\sin(\l_2+\xi-u-\eta)\sin
2u}
{\sin(\l_2+\bar\xi-u-\eta)\sin(\l_1+\xi+u)\sin(\l_2+\xi+u)\sin(2u+\eta)}\no\\
&&~~~~~~~~~~~~~~~~~~\times\prod_{k=1}^M\frac{\sin(u+v^{(1)}_k+2\eta)\sin(u-v^{(1)}_k+\eta)}
{\sin(u+v^{(1)}_k+\eta)\sin(u-v^{(1)}_k)}\no\\
&&~~~~~~~~~~~~~~~~~~\times\prod_{k=1}^{2M}\frac{\sin(u+z_k)\sin(u-z_k)}
{\sin(u+z_k+\eta)\sin(u-z_k+\eta)}.\label{Eigenfuction-D-1}
 \eea

\noindent If the parameters $\{v^{(2)}_k\}$ satisfy the second
Bethe Ansatz equations \bea
&&\hspace{-0.1cm}\frac
  {\sin(\l_1+\xi+v^{(2)}_{\a})\sin(\l_1+\bar\xi-v^{(2)}_{\a})
  \sin(\l_2+\bar\xi+v^{(2)}_{\a})\sin(\l_2+\xi-v^{(2)}_{\a})}
  {\sin(\l_1\hspace{-0.1cm}+\hspace{-0.1cm}
  \bar\xi\hspace{-0.1cm}+\hspace{-0.1cm}v^{(2)}_{\a}
  \hspace{-0.1cm}+\hspace{-0.1cm}\eta)
  \sin(\l_1\hspace{-0.1cm}+\hspace{-0.1cm}\xi\hspace{-0.1cm}-\hspace{-0.1cm}v^{(2)}_{\a}
  \hspace{-0.1cm}-\hspace{-0.1cm}\eta)
  \sin(\l_2\hspace{-0.1cm}+\hspace{-0.1cm}\xi\hspace{-0.1cm}+\hspace{-0.1cm}
  v^{(2)}_{\a}\hspace{-0.1cm}+\hspace{-0.1cm}\eta)
  \sin(\l_2\hspace{-0.1cm}+\hspace{-0.1cm}\bar\xi\hspace{-0.1cm}-\hspace{-0.1cm}v^{(2)}_{\a}
  \hspace{-0.1cm}-\hspace{-0.1cm}\eta)}\no\\
&&~~~~~~=\prod_{k\neq
  \a}^M\frac{\sin(v^{(2)}_{\a}+v^{(2)}_k+2\eta)\sin(v^{(2)}_{\a}-v^{(2)}_k+\eta)}
  {\sin(v^{(2)}_{\a}+v^{(2)}_k)\sin(v^{(2)}_{\a}-v^{(2)}_k-\eta)}\no\\
&&~~~~~~~~~~\times\prod_{k=1}^{2M}\frac{\sin(v^{(2)}_{\a}+z_k)\sin(v^{(2)}_{\a}-z_k)}
  {\sin(v^{(2)}_{\a}+z_k+\eta)\sin(v^{(2)}_{\a}-z_k+\eta)},~~\a=1,\cdots,M,
  \label{BA-D-2}
\eea the Bethe states $|v^{(2)}_1,\cdots,v^{(2)}_M\rangle^{(II)}$
yield the second set of the eigenstates of the transfer matrix
with the eigenvalues \cite{Yan07}, \bea
&&\L^{(2)}(u)=\frac{\sin(2u+2\eta)\sin(\l_1+\bar\xi-u)\sin(\l_2+\bar\xi+u)\sin(\l_2+\xi-u)}
{\sin(2u+\eta)\sin(\l_1+\bar\xi-u-\eta)\sin(\l_2+\bar\xi-u-\eta)\sin(\l_2+\xi+u)}\no\\
&&~~~~~~~~~~~~~~~~~~\times\prod_{k=1}^M\frac{\sin(u+v^{(2)}_k)\sin(u-v^{(2)}_k-\eta)}
{\sin(u+v^{(2)}_k+\eta)\sin(u-v^{(2)}_k)}\no\\
&&~~~~~~+\frac{\sin(2u)\sin(\l_1+\bar\xi+u+\eta)
\sin(\l_2+\xi+u+\eta)\sin(\l_1+\xi-u-\eta)}
{\sin(2u+\eta)\sin(\l_1+\bar\xi-u-\eta)\sin(\l_2+\xi+u)\sin(\l_1+\xi+u)}\no\\
&&~~~~~~~~~~~~~~~~~~\times\prod_{k=1}^M\frac{\sin(u+v^{(2)}_k+2\eta)\sin(u-v^{(2)}_k+\eta)}
{\sin(u+v^{(2)}_k+\eta)\sin(u-v^{(2)}_k)}\no\\
&&~~~~~~~~~~~~~~~~~~\times\prod_{k=1}^{2M}\frac{\sin(u+z_k)\sin(u-z_k)}
{\sin(u+z_k+\eta)\sin(u-z_k+\eta)}.\label{Eigenfuction-D-2}
 \eea


\section{ A new set of  Bethe states}
\label{NE} \setcounter{equation}{0}

In contrast with the case of periodic boundary conditions
\cite{Mai00}, we find that in the F-basis $\T^-(m|u)^1_2$ and
$\T^-(m|u)^2_1$ given in (\ref{Mon-F}) cannot {\it simultaneously}
be expressed in simple polarization free form. A similar situation
already appeared in the open XXZ chain with diagonal K-matrices
\cite{Wan02,Kit07}. However, such polarization free forms of the
pseudo-particle creation operators are crucial for computing
partition functions and scalar products
\cite{Kit99,Yan06-1,Zha06,Kit07,Yan09}.

To overcome the difficulty, let us introduce another double-row
monodromy matrix $\T^+(m|u)$ (c.f.  $\T^-(m|u)$ given by
(\ref{Mon-F})) \bea
 \T^+(m|u)^j_i&=&\prod_{k\neq j}\frac{\sin(m_{jk})}{\sin(m_{jk}-\eta)}
      \,\phi^{t_0}_{m-\eta(\hat{\jmath}-\hat{\imath}),m-\eta\hat{\jmath}}(u)
      \lt(\mathbb{T}^+(u)\rt)^{t_0}\bar{\phi}^{t_0}_{m,m-\eta\hat{\jmath}}(-u),
      \label{Mon-F-1}\\
 \lt(\mathbb{T}^+(u)\rt)^{t_0}&=&T^{t_0}(u)\lt(K^+(u)\rt)^{t_0}\hat{T}^{t_0}(u),
      \label{Mon-V-1}
\eea where $t_0$ denotes transposition in the $0$-th space (i.e.
auxiliary space). As will be shown in section 5, in the F-basis
the face versions of the operators $\T^+(m|u)^1_2$ and
$\T^-(m|u)^2_1$ become completely symmetric and polarization free
simultaneously (see (\ref{Creation-operator-1}) and
(\ref{Creation-operator-2}) below). Thus $\T^+(m|u)^1_2$ is the
desirable pseudo-particle creation  operator which generates the
new set of Bethe states $|v^{(1)}_1,\cdots,v^{(1)}_M\rangle^{(I)}
$, in replacement of the first set of Bethe states
(\ref{Bethe-state}), \bea
&&|v^{(1)}_1,\cdots,v^{(1)}_M\rangle^{(I)}=
     \T^+(\l+2\eta\hat{1}|v^{(1)}_1)^1_2\cdots
   \T^+(\l+2M\eta\hat{1}|v^{(1)}_M)^1_2|\O^{(I)}(\l)\rangle
   \label{Bethe-state-1}
\eea with reference state $|\O^{(I)}(\l)\rangle$ (c.f.
$|\O^{(1)}(\l)\rangle$ given by (\ref{Vac}))
\bea
\hspace{-1.2truecm}|\O^{(I)}(\l)\rangle
   &=&\phi^1_{\l+N\eta\hat{1},\l+(N-1)\eta\hat{1}}(z_1)
      \phi^2_{\l+(N-1)\eta\hat{1},\l+(N-2)\eta\hat{1}}(z_{2})\cdots
      \phi^N_{\l+\eta\hat{1},\l}(z_N).\label{Vac-1}
\eea As will be seen in section 5, the two sets of Bethe states
(\ref{Bethe-state-1}) and (\ref{Bethe-state-2}) are of completely
symmetric form in the F-basis. Moreover,  using the technique
developed in \cite{Yan04}, we find that if the parameters
$\{v^{(1)}_k\}$ in (\ref{Bethe-state-1}) satisfy the first set of
Bethe ansatz equations (\ref{BA-D-1}), the state
$|v^{(1)}_1,\cdots,v^{(1)}_M\rangle^{(I)}$ is the eigenstate of
the transfer matrix with eigenvalue $\L^{(1)}(u)$ given by
(\ref{Eigenfuction-D-1}) (for details, see Appendix A). Hence two
sets of Bethe states (\ref{Bethe-state-1}) and
(\ref{Bethe-state-2}) constitute a complete set.


\section{ $\T^{\pm}(m|u)$ in the face picture}
\label{T} \setcounter{equation}{0}

The K-matrices $K^{\pm}(u)$ given by (\ref{K-matrix}) and
(\ref{DK-matrix}) are generally non-diagonal (in the vertex
picture), after the face-vertex transformations (\ref{K-F-1}) and
(\ref{K-F-2}), the face type counterparts $\K(\l|u)$ and
$\tilde{\K}(\l|u)$ given by (\ref{K-F-3}) and (\ref{K-F-4}) {\it
simultaneously\/} become diagonal. This fact suggests that it
would be much simpler if one performs all calculations in the face
picture.

Let us introduce the face type one-row monodromy matrix (c.f
(\ref{Mon-V})) \bea
 T_{F}(l|u)&\equiv &T^{F}_{0,1\ldots N}(l|u)\no\\
 &=&R_{0,N}(u-z_N;l-\eta\sum_{i=1}^{N-1}h^{(i)})\ldots
    R_{0,2}(u-z_2;l-\eta h^{(1)})R_{0,1}(u-z_1;l),\no\\
 &=&\lt(\begin{array}{ll}T_F(l|u)^1_1&T_F(l|u)^1_2\\T_F(l|u)^2_1&
   T_F(l|u)^2_2\end{array}\rt)
    \label{Monodromy-face-1}
\eea where $l$ is a generic vector in $V$. The monodromy matrix
satisfies the face type quadratic exchange relation
\cite{Fel96,Hou03}. Applying $T_F(l|u)^i_j$ to an arbitrary vector
$|i_1,\ldots,i_N\rangle$ in the N-tensor product space $V^{\otimes
N}$ given by \bea
   |i_1,\ldots,i_N\rangle=\e^1_{i_1}\ldots
   \e^N_{i_N},\label{Vector-V}
\eea we have \bea
 T_F(l|u)^i_j|i_1,\ldots,i_N\rangle&\equiv&
    T_F(m;l|u)^i_j|i_1,\ldots,i_N\rangle\no\\
 &=&\sum_{\a_{N-1}\ldots\a_1}\sum_{i'_N\ldots i'_1}
 R(u-z_N;l-\eta\sum_{k=1}^{N-1}\hat{\imath}'_k)
   ^{i\,\,\,\,\,\,\,\,\,\,\,\,\,\,i'_N}_{\a_{N-1}\,i_N}\ldots\no\\
 &&\quad\quad\times R(u-z_2;l-\eta\hat{\imath}'_1)^{\a_2\,i'_2}_{\a_1\,\,i_2}
 R(u-z_1;l)^{\a_1\,i'_1}_{j\,\,\,\,i_1}
   \,\,|i'_1,\ldots,i'_N\rangle,\label{Monodromy-face-2}
\eea where $m=l-\eta\sum_{k=1}^N\hat{\imath}_k$. We shall express
the double-row monodromy matrices $\T^{\pm}$ given by
(\ref{Mon-F}) and (\ref{Mon-F-1}) in terms of the above face-type
one-row monodromy matrix.

Associated with the vertex type monodromy matrices $T(u)$
(\ref{Mon-V}) and $\hat{T}(u)$ (\ref{Mon-V-0}), we introduce the
following operators \bea
 T(m,l|u)^j_{\mu}&=&\tilde{\phi}^0_{m+\eta\hat{\jmath},m}(u)\,T_0(u)\,
    \phi^0_{l+\eta\hat{\mu},l}(u),\\
 S(m,l|u)^{\mu}_{i}&=&\bar{\phi}^0_{l,l-\eta\hat{\mu}}(-u)\,\hat{T}_0(u)\,
    \phi^0_{m,m-\eta\hat{\imath}}(-u).
\eea Moreover, for the case of
$m=l-\eta\sum_{k=1}^N\hat{\imath}_k$, we introduce a generic state
in the quantum space from  the intertwiner vector (\ref{Intvect})
\bea
 |i_1,\ldots,i_N\rangle^{m}_{l}=
     \phi^1_{l,l-\eta\hat{\imath}_1}(z_1)
     \phi^2_{l-\eta\hat{\imath}_1,l-\eta(\hat{\imath}_1+\hat{\imath}_2)}(z_2)\ldots
     \phi^N_{l-\eta\sum_{k=1}^{N-1}\hat{\imath}_k,l-\eta\sum_{k=1}^{N}\hat{\imath}_k}(z_N).
\eea We can evaluate the action of the operator $T(m,l|u)$ on the
state $|i_1,\ldots,i_N\rangle^{m}_{l}$ from the face-vertex
correspondence relation (\ref{Face-vertex}) \bea
 &&T(m,l|u)^j_{\mu}|i_1,\ldots,i_N\rangle^{m}_{l}=
    \tilde{\phi}^0_{m+\eta\hat{\jmath},m}(u)\,T_0(u)\,
    \phi^0_{l+\eta\hat{\mu},l}(u)|i_1,\ldots,i_N\rangle^{m}_{l}
    \no\\
 &&\quad\quad=\tilde{\phi}^0_{m+\eta\hat{\jmath},m}(u)
    \R_{0,N}(u-z_N)\ldots\R_{0,1}(u-z_1)\phi^0_{l+\eta\hat{\mu},l}(u)
    \phi^1_{l,l-\eta\hat{\imath}_1}(z_1)\ldots\no\\
 &&\quad\quad=\sum_{\a_1,i'_1}R(u-z_1;l+\eta\hat{\mu})^{\a_1i'_1}_{\mu\,\, i_1}
   \phi^1_{l+\eta\hat{\mu},l+\eta\hat{\mu}-\eta\hat{\imath}'_1}(z_1)
   \tilde{\phi}^0_{m+\eta\hat{\jmath},m}(u)
    \R_{0,N}(u-z_N)\ldots\no\\
 &&\quad\quad\quad\quad \times
   \R_{0,2}(u-z_2)\phi^0_{l+\eta\hat{\mu}-\eta\hat{\imath}'_1,l-\eta\hat{\imath}_1}(u)
    \phi^2_{l-\eta\hat{\imath}_1,l-\eta(\hat{\imath}_1+\hat{\imath}_2)}(z_2)\ldots\no\\
 &&\quad\quad\vdots\no\\
 &&\quad\quad=\sum_{\a_{1}\ldots\a_{N-1}}\sum_{i'_1\ldots i'_N}
    R(u-z_N;l+\eta\hat{\mu}-\eta\sum_{k=1}^{N-1}\hat{\imath}'_k)
    ^{j\,\,\,\,\,\,\,\,\,\,\,\,i'_N}_{\a_{N-1}i_N}\ldots\no\\
 &&\quad\quad\quad\quad\times R(u-z_1;l+\eta\hat{\mu})^{\a_1i'_1}_{\mu \,\,i_1}
    |i'_1,\ldots,i'_N\rangle^{l+\eta\hat{\mu}-\eta\sum_{k=1}^N\hat{\imath}'_k}_{l+\eta\hat{\mu}}.
\eea  Here we have used the following property of the R-matrix
\bea
  R(u;m)^{i'j'}_{ij}=R(u;m\pm\eta(\hat{\imath}+\hat{\jmath}))^{i'j'}_{ij}
  =R(u;m\pm\eta(\hat{\imath}'+\hat{\jmath}'))^{i'j'}_{ij},
\eea and the weight conservation condition (\ref{Conservation}).
Comparing with (\ref{Monodromy-face-2}), we have the following
correspondence \bea
 T(m,l|u)^j_{\mu}|i_1,\ldots,i_N\rangle^{m}_{l}\,\longleftrightarrow\,
      T_F(m+\eta\hat{\mu};l+\eta\hat{\mu}|u)_{\mu}^j|i_1,\ldots,i_N\rangle,\label{crospendence-1}
\eea where vector $|i_1,\ldots,i_N\rangle$ is given by
(\ref{Vector-V}). Hereafter, we will use  $O_F$ to denote the face
version of operator $O$ in the face picture.

Noting that \bea
 \hat{T}_0(u)=\R_{1,0}(u+z_1)\ldots\R_{N,0}(u+z_N),\no
\eea we obtain the action of $S(m,l|u)^{\mu}_{i}$ on the state
$|i_1,\ldots,i_N\rangle^{m}_{l}$ \bea
 S(m,l|u)^{\mu}_{i}|i_1,\ldots,i_N\rangle^{m}_{l}
     \hspace{-0.22truecm}&=&\hspace{-0.42truecm}
     \sum_{\a_{1}\ldots\a_{N-1}}\sum_{i'_1\ldots i'_N}
     R(u+z_1;l)^{i'_1\,\mu}_{i_1\a_{N-1}}
     R(u+z_2;l-\eta\hat{\imath}_1)^{i'_2\,\a_{N-1}}_{i_2\a_{N-2}}\no\\
 &&\,\times
     \ldots R(u\hspace{-0.12truecm}+\hspace{-0.12truecm}z_N;
     l\hspace{-0.12truecm}-\hspace{-0.12truecm}\eta
     \sum_{k=1}^{N-1}\hat{\imath}_k)
     ^{i'_N\,\a_1}_{i_N\,i}|i'_1,\ldots,i'_N\rangle
     ^{l-\eta\hat{\mu}-\eta\sum_{k=1}^N\hat{\imath}'_k}_{l-\eta\hat{\mu}}.\no\\
\eea Then the crossing relation of the R-matrix (\ref{Crossing})
enables us to establish the following relation:
\bea
  S(m,l|u)^{\mu}_{i}=\varepsilon_{\bar{i}}\varepsilon_{\bar{\mu}}
   \frac{\sin\lt(m_{21}\rt)}{\sin\lt(l_{21}\rt)}
   \prod_{k=1}^N\frac{\sin(u+z_k)}{\sin(u+z_k+\eta)}
   T(m,l|-u-\eta)^{\bar{i}}_{\bar{\mu}},
   \label{Crossing-operator}
\eea where the parities are defined in (\ref{Parity}) and $m_{21}$
(or $l_{21}$) is defined in (\ref{Def1}).

Now we are in the position to express $\T^{\pm}$ (\ref{Mon-F}) and
(\ref{Mon-F-1}) in terms of $T(m,l)^i_j$ and $S(l,m)^i_j$. By
(\ref{Int3}) and (\ref{Int4}),  we have \bea
  \T^-(m|u)^j_i&=&\tilde{\phi}^{0}_{m-\eta(\hat{\imath}-\hat{\jmath}),
      m-\eta\hat{\imath}}(u)~\mathbb{T}(u)~\phi^{0}_{m,
      m-\eta\hat{\imath}}(-u)\no\\
  &=&\tilde{\phi}^{0}_{m-\eta(\hat{\imath}-\hat{\jmath}),
      m-\eta\hat{\imath}}(u)T_0(u)K^-_0(u)\hat{T}_0(u)\phi^{0}_{m,
      m-\eta\hat{\imath}}(-u)\no\\
  &=&\hspace{-0.32truecm}\sum_{\mu,\nu}\tilde{\phi}^{0}_{m-\eta(\hat{\imath}-\hat{\jmath}),
      m-\eta\hat{\imath}}(u)T_0(u)
      \phi^0_{l-\eta(\hat{\nu}-\hat{\mu}),l-\eta\hat{\nu}}(u)
      \tilde{\phi}^0_{l-\eta(\hat{\nu}-\hat{\mu}),l-\eta\hat{\nu}}(u)\no\\
  &&\quad\times K^-_0(u)\phi^0_{l,l-\eta\hat{\nu}}(-u)
      \bar{\phi}^0_{l,l-\eta\hat{\nu}}(-u)
      \hat{T}_0(u)\phi^{0}_{m,
      m-\eta\hat{\imath}}(-u)\no\\
  &=&\hspace{-0.22truecm}\sum_{\mu,\nu}T(m-\eta\hat{\imath},l-\eta\hat{\nu}|u)^j_{\mu}
       \K(l|u)^{\mu}_{\nu}S(m,l|u)_i^{{\nu}}\no\\
  &\stackrel{{\rm def}}{=}& \T^-(m,l|u)^j_i,
\eea where the face-type K-matrix $\K(l|u)^{\mu}_{\nu}$ is given
by
\bea
 \K(l|u)^{\mu}_{\nu}=\tilde{\phi}^0_{l-\eta(\hat{\nu}-\hat{\mu}),l-\eta\hat{\nu}}(u)
    K^-_0(u)\phi^0_{l,l-\eta\hat{\nu}}(-u).\label{K-1}
\eea Similarly, we have
\bea
 \T^+(m|u)^j_i&=&\prod_{k\neq j}\frac{\sin m_{jk}}{\sin\lt(m_{jk}-\eta\rt)}
       \sum_{\mu,\nu}T(l-\eta\hat{\mu},m-\eta\hat{\jmath}|u)_i^{\nu}
       \tilde{\K}(l|u)^{\mu}_{\nu}S(l,m|u)^j_{{\mu}}\no\\
&\stackrel{{\rm
       def}}{=}&\T^+(l,m|u)^j_i
\eea with
\bea
 \tilde{\K}(l|u)^{\mu}_{\nu}=\bar{\phi}^0_{l,l-\eta\hat{\mu}}(-u)
    K^+_0(u)\phi^0_{l-\eta(\hat{\mu}-\hat{\nu}),l-\eta\hat{\mu}}(u).
    \label{K-2}
\eea Thanks to the fact that when $l=\l$ the corresponding
face-type K-matrices $\K(\l|u)$  (\ref{K-1}) and
$\tilde{\K}(\l|u)$  (\ref{K-2}) become diagonal ones (\ref{K-F-3})
and (\ref{K-F-4}), we have \bea
 \T^-(m,\l|u)^j_i\hspace{-0.22truecm}&=&\hspace{-0.22truecm}
       \sum_{\mu}T(m-\eta\hat{\imath},\l-\eta\hat{\mu}|u)^j_{\mu}
       k(\l|u)_{\mu}S(m,\l|u)_i^{{\mu}},\label{New-M-F-1}\\
 \T^+(\l,m|u)^j_i\hspace{-0.22truecm}&=&\hspace{-0.22truecm}
\prod_{k\neq j}\frac{\sin m_{jk}}{\sin\lt(m_{jk}-\eta\rt)}
       \sum_{\mu}T(\l-\eta\hat{\mu},m-\eta\hat{\jmath}|u)_i^{\mu}
       \tilde{k}(\l|u)_{\mu}S(\l,m|u)^j_{{\mu}},\label{New-M-F-2}
 \eea where the functions $k(\l|u)_{\mu}$ and $\tilde{k}(\l|u)_{\mu}$ are given by
 (\ref{K-F-3}) and (\ref{K-F-4}) respectively. The relation
(\ref{Crossing-operator}) implies that one can further express
$\T^{\pm}(m|u)^j_i$ in terms of only $T(m,l|u)^j_i$. Here we
present the results for the pseudo-particle creation operators
$\T^{-}(m|u)^2_1$ in (\ref{Bethe-state-2}) and $\T^{+}(m|u)^1_2$
in (\ref{Bethe-state-1}): \bea
 \T^-(m|u)^2_1\hspace{-0.22truecm}&=&\hspace{-0.22truecm}
      \T^-(m,\l|u)^2_1=\frac{\sin(m_{21})}{\sin(\l_{21})}\prod_{k=1}^N
      \frac{\sin(u+z_k)}{\sin(u+z_k+\eta)}\no\\
      &&\,\times\lt\{
      \frac{\sin(\l_1+\xi-u)}{\sin(\l_1+\xi+u)}
      T(m+\eta\hat{2},\l+\eta\hat{2}|u)^2_1
      T(m,\l|-u-\eta)^2_2\rt.\no\\
 &&\,\quad-\lt.
      \frac{\sin(\l_2+\xi-u)}{\sin(\l_2+\xi+u)}
      T(m+\eta\hat{2},\l+\eta\hat{1}|u)^2_2
      T(m,\l|-u-\eta)^2_1\rt\},\label{Expression-1}\\
 \T^+(m|u)^1_2\hspace{-0.22truecm}&=&\hspace{-0.22truecm}
      \T^+(\l,m|u)^1_2=\prod_{k=1}^N
      \frac{\sin(u+z_k)}{\sin(u+z_k+\eta)}\no\\
      &&\,\times\lt\{
      \frac{\sin(\l_{12}\hspace{-0.08truecm}-\hspace{-0.08truecm}\eta)
      \sin(\l_1\hspace{-0.08truecm}+\hspace{-0.08truecm}\bar{\xi}\hspace{-0.08truecm}+\hspace{-0.08truecm}u
      \hspace{-0.08truecm}+\hspace{-0.08truecm}\eta)}
      {\sin(m_{12}\hspace{-0.12truecm}-\hspace{-0.12truecm}\eta)
      \sin(\l_1\hspace{-0.12truecm}+\hspace{-0.12truecm}\bar{\xi}\hspace{-0.12truecm}-\hspace{-0.12truecm}u
      \hspace{-0.12truecm}-\hspace{-0.12truecm}\eta)}
      T(\l\hspace{-0.12truecm}+\hspace{-0.12truecm}\eta\hat{2},m\hspace{-0.12truecm}+\hspace{-0.12truecm}\eta\hat{2}|u)^1_2
      T(\l,m|\hspace{-0.12truecm}-\hspace{-0.12truecm}u\hspace{-0.12truecm}-\hspace{-0.12truecm}\eta)^2_2\rt.\no\\
 &&\,\quad\hspace{-0.12truecm}-\hspace{-0.12truecm}\lt.
      \frac{\sin(\l_{21}\hspace{-0.12truecm}-\hspace{-0.12truecm}\eta)
      \sin(\l_2\hspace{-0.12truecm}+\hspace{-0.12truecm}\bar{\xi}\hspace{-0.12truecm}+\hspace{-0.12truecm}u
      \hspace{-0.12truecm}+\hspace{-0.12truecm}\eta)}
      {\sin(m_{21}\hspace{-0.12truecm}+\hspace{-0.12truecm}\eta)
      \sin(\l_2\hspace{-0.12truecm}+\hspace{-0.12truecm}\bar{\xi}\hspace{-0.12truecm}-\hspace{-0.12truecm}u
      \hspace{-0.12truecm}-\hspace{-0.12truecm}\eta)}
      T(\l\hspace{-0.12truecm}+\hspace{-0.12truecm}\eta\hat{1},m
      \hspace{-0.12truecm}+\hspace{-0.12truecm}\eta\hat{2}|u)^2_2
      T(\l,m|\hspace{-0.12truecm}-\hspace{-0.12truecm}u\hspace{-0.12truecm}
      -\hspace{-0.12truecm}\eta)^1_2\rt\}.\no\\
 &&\label{Expression-2}
\eea Similar to (\ref{crospendence-1}), we have the
correspondence,
\bea
 &&\T^-(m,l|u)^2_1|i_1,\ldots,i_N\rangle^{m}_{l}\,\longleftrightarrow\,
   \T^-_{F}(m,l|u)^2_1|i_1,\ldots,i_N\rangle,\\
 &&\T^+(m,l|u)^1_2|i_1,\ldots,i_N\rangle^{m}_{l}\,\longleftrightarrow\,
   \T^+_{F}(m,l|u)^1_2|i_1,\ldots,i_N\rangle.
\eea It follows from (\ref{Expression-1}) and (\ref{Expression-2})
that the face-type double-row monodromy matrix elements
$\T^-_F(m,\l|u)^2_1$ and $\T^+_F(\l,m|u)^1_2$ can be expressed in
terms of the face-type one-row monodromy matrix elements
$T_F(m,l|u)^i_j $ (\ref{Monodromy-face-2}) by \bea
 &&\T^-_F(m,\l|u)^2_1=
 \frac{\sin(m_{21})}{\sin(\l_{21})}\prod_{k=1}^N
      \frac{\sin(u+z_k)}{\sin(u+z_k+\eta)}\no\\
 &&\,\quad\times\lt\{
      \frac{\sin(\l_1+\xi-u)}{\sin(\l_1+\xi+u)}
      T_F(m,\l|u)^2_1
      T_F(m+\eta\hat{2},\l+\eta\hat{2}|-u-\eta)^2_2\rt.\no\\
 &&\,\qquad-\lt.
      \frac{\sin(\l_2+\xi-u)}{\sin(\l_2+\xi+u)}
      T_F(m+2\eta\hat{2},\l|u)^2_2
      T_F(m+\eta\hat{1},\l+\eta\hat{1}|-u-\eta)^2_1\rt\},\label{Expression-3}\\
 &&\T^+_F(\l,m|u)^1_2=\prod_{k=1}^N
      \frac{\sin(u+z_k)}{\sin(u+z_k+\eta)}\no\\
 &&\,\quad\times\lt\{
      \frac{\sin(\l_{12}\hspace{-0.08truecm}-\hspace{-0.08truecm}\eta)
      \sin(\l_1\hspace{-0.08truecm}+\hspace{-0.08truecm}\bar{\xi}
      \hspace{-0.08truecm}+\hspace{-0.08truecm}u\hspace{-0.08truecm}+\hspace{-0.08truecm}\eta)}
      {\sin(m_{12}\hspace{-0.08truecm}-\hspace{-0.08truecm}\eta)
      \sin(\l_1\hspace{-0.08truecm}+\hspace{-0.08truecm}\bar{\xi}\hspace{-0.08truecm}-
      \hspace{-0.08truecm}u\hspace{-0.08truecm}-\hspace{-0.08truecm}\eta)}
      T_F(\l\hspace{-0.08truecm}+\hspace{-0.08truecm}2\eta\hat{2},m
      \hspace{-0.08truecm}+\hspace{-0.08truecm}2\eta\hat{2}|u)^1_2
      T_F(\l\hspace{-0.08truecm}+\hspace{-0.08truecm}\eta\hat{2},m
      \hspace{-0.08truecm}+\hspace{-0.08truecm}\eta\hat{2}|
      \hspace{-0.08truecm}-\hspace{-0.08truecm}u
      \hspace{-0.08truecm}-\hspace{-0.08truecm}\eta)^2_2\rt.\no\\
 &&\,\qquad-\lt.
      \frac{\sin(\l_{21}\hspace{-0.08truecm}-\hspace{-0.08truecm}\eta)
      \sin(\l_2\hspace{-0.08truecm}+\hspace{-0.08truecm}\bar{\xi}
      \hspace{-0.08truecm}+\hspace{-0.08truecm}u
      \hspace{-0.08truecm}+\hspace{-0.08truecm}\eta)}
      {\sin(m_{21}\hspace{-0.08truecm}+\hspace{-0.08truecm}\eta)
      \sin(\l_2\hspace{-0.08truecm}+\hspace{-0.08truecm}\bar{\xi}
      \hspace{-0.08truecm}-\hspace{-0.08truecm}u
      \hspace{-0.08truecm}-\hspace{-0.08truecm}\eta)}
      T_F(\l,m\hspace{-0.08truecm}+\hspace{-0.08truecm}2\eta\hat{2}|u)^2_2
      T_F(\l\hspace{-0.08truecm}+\hspace{-0.08truecm}\eta\hat{2},m
      \hspace{-0.08truecm}+\hspace{-0.08truecm}\eta\hat{2}|
      \hspace{-0.08truecm}-\hspace{-0.08truecm}u\hspace{-0.08truecm}-\hspace{-0.08truecm}
      \eta)^1_2\rt\}.\no\\
 &&\label{Expression-4}
\eea In the derivation of the above equations we have used the
identity $\hat{1}+\hat{2}=0$. Finally, we obtain the face versions
$|v^{(1)}_1,\cdots,v^{(1)}_M\rangle^{(I)}_F$ and
$|v^{(1)}_1,\cdots,v^{(1)}_M\rangle^{(II)}_F$ of the two sets of
Bethe states (\ref{Bethe-state-1}) and (\ref{Bethe-state-2}),
\bea
 &&|v^{(1)}_1,\cdots,v^{(1)}_M\rangle^{(I)}_F=
     \T^+_F(\l,\l\hspace{-0.08truecm}+\hspace{-0.08truecm}2\eta\hat{1}
     |v^{(1)}_1)^1_2\cdots
   \T^+_F(\l,\l\hspace{-0.08truecm}+\hspace{-0.08truecm}2M\eta\hat{1}
   |v^{(1)}_M)^1_2|1,\ldots,1\rangle,
   \label{Bethe-state-3}\\
 &&|v^{(2)}_1,\cdots,v^{(2)}_M\rangle^{(II)}_F =
   \T^-_F(\l\hspace{-0.08truecm}-\hspace{-0.08truecm}2\eta\hat{2},\l
   |v^{(2)}_1)^2_1
   \cdots
   \T^-_F(\l\hspace{-0.08truecm}-\hspace{-0.08truecm}2M\eta\hat{2},\l
   |v^{(2)}_M)^2_1
   |2,\ldots,2\rangle.
   \label{Bethe-state-4}
\eea

In the next section we shall construct the Drinfeld twist (or
factorizing F-matrix) in the face picture for the open XXZ chain
with non-diagonal boundary terms. In this F-basis, the two sets of
pseudo-particle creation operators $\T^{\pm}_F$ given by
(\ref{Expression-3}) and (\ref{Expression-4}) take  completely
symmetric and polarization free forms simultaneously. Moreover,
the corresponding two sets of Bethe states become completely
symmetric.


\section{ F-basis}
\label{F} \setcounter{equation}{0}

In this section, we construct the Drinfeld twist \cite{Dri83}
(factorizing F-matrix) on the $N$-fold tensor product space
$V^{\otimes N}$ (i.e. the quantum space of the open XXZ chain) and
the associated representations of the pseudo-particle creation
operators in this basis.

\subsection{Factorizing Drinfeld twist $F$}
Let $ \mathcal{S}_N$ be the permutation group over indices
$1,\ldots,N$ and $\{\s_i|i=1,\ldots,N-1\}$ be the set of
elementary permutations in $\mathcal{S}_N$. For each elementary
permutation $\s_i$, we introduce the associated operator
$R^{\s_i}_{1\ldots N}$ on the quantum space \bea
  R^{\s_i}_{1\ldots N}(l)\equiv R^{\s_i}(l)=R_{i,i+1}
    (z_i-z_{i+1}|l-\eta\sum_{k=1}^{i-1}h^{(k)}),\label{Fundamental-R-operator}
\eea where $l$ is a generic vector in $V$. For any $\s,\,\s'\in
\mathcal{S}_N$, operator $R^{\s\s'}_{1\ldots N}$ associated with
$\s\s'$ satisfies the following composition law
 \cite{Mai00,Alb00,Alb00-1,Yan06-1}:
\bea
  R_{1\ldots N}^{\s\s'}(l)=R^{\s'}_{\s(1\ldots
  N)}(l)\,R^{\s}_{1\ldots N}(l).\label{Rule}
\eea Let $\s$ be decomposed in a minimal way in terms of
elementary permutations,
\bea
  \s=\s_{\b_1}\ldots\s_{\b_p}, \label{decomposition}
\eea where $\b_i=1,\ldots, N-1$ and the positive integer $p$ is
the length of $\s$. The composition law (\ref{Rule}) enables one
to obtain  operator $R^{\s}_{1\ldots N}$ associated with each
$\s\in\mathcal{S}_N $. The dynamical quantum Yang-Baxter equation
(\ref{MYBE}), weight conservation condition (\ref{Conservation})
and unitary condition (\ref{Unitary}) guarantee the uniqueness of
$R^{\s}_{1\ldots N}$. Moreover, one may check that
$R^{\s}_{1\ldots N}$ satisfies the following exchange relation
with the face type one-row monodromy matrix
(\ref{Monodromy-face-1}) \bea
  R^{\s}_{1\ldots N}(l)T^F_{0,1\ldots N}(l|u)=T^F_{0,\s(1\ldots N)}(l|u)
    R^{\s}_{1\ldots N}(l-\eta h^{(0)}),\quad\quad \forall\s\in
    \mathcal{S}_N.\label{Exchang-Face-1}
\eea

Now, we construct the face-type Drinfeld twist $F_{1\ldots
N}(l)\equiv F_{1\ldots N}(l;z_1,\ldots,z_N)$ \footnote{In this
paper, we adopt the convention: $F_{\s(1\ldots N)}(l)\equiv
F_{\s(1\ldots N)}(l;z_{\s(1)},\ldots,z_{\s(N)})$.} on the $N$-fold
tensor product space $V^{\otimes N}$, which  satisfies the
following three properties \cite{Alb00,Yan06-1}: \bea
 &&{\rm I.\,\,\,\,lower-triangularity;}\\
 &&{\rm II.\,\,\, non-degeneracy;}\\
 &&{\rm III.\,factorizing \, property}:\,\,
 R^{\s}_{1\ldots N}(l)\hspace{-0.08truecm}=\hspace{-0.08truecm}
    F^{-1}_{\s(1\ldots N)}(l)F_{1\ldots N}(l), \,\,
 \forall\s\in  \mathcal{S}_N.\label{Factorizing}
\eea Substituting (\ref{Factorizing}) into the exchange relation
(\ref{Exchang-Face-1}), we have
\bea
 F^{-1}_{\s(1\ldots N)}(l)F_{1\ldots N}(l)T^F_{0,1\ldots N}(l|u)=
   T^F_{0,\s(1\ldots N)}(l|u)F^{-1}_{\s(1\ldots N)}(l-\eta h^{(0)})
   F_{1\ldots N}(l-\eta h^{(0)}).
\eea Equivalently,
\bea
 F_{1\ldots N}(l)T^F_{0,1\ldots N}(l|u)F^{-1}_{1\ldots N}(l-\eta h^{(0)})
   =F_{\s(1\ldots N)}(l)T^F_{0,\s(1\ldots N)}(l|u)
   F^{-1}_{\s(1\ldots N)}(l-\eta h^{(0)}).\label{Invariant}
\eea Let us introduce the twisted monodromy matrix
$\tilde{T}^F_{0,1\ldots N}(l|u)$ by \bea
 \tilde{T}^F_{0,1\ldots N}(l|u)&=&
  F_{1\ldots N}(l)T^F_{0,1\ldots N}(l|u)F^{-1}_{1\ldots N}(l-\eta
  h^{(0)})\no\\
  &=&\lt(\begin{array}{ll}\tilde{T}_F(l|u)^1_1&\tilde{T}_F(l|u)^1_2
  \\\tilde{T}_F(l|u)^2_1&
   \tilde{T}_F(l|u)^2_2\end{array}\rt).\label{Twisted-Mon-F}
\eea Then (\ref{Invariant}) implies that the twisted monodromy
matrix is symmetric under $\mathcal{S}_N$, namely, \bea
 \tilde{T}^F_{0,1\ldots N}(l|u)=\tilde{T}^F_{0,\s(1\ldots
 N)}(l|u), \quad \forall \s\in \mathcal{S}_N.
\eea

Define the F-matrix:
\bea
  F_{1\ldots N}(l)=\sum_{\s\in
     \mathcal{S}_N}\sum^2_{\{\a_j\}=1}\hspace{-0.22truecm}{}^*\,\,\,\,
     \prod_{i=1}^NP^{\s(i)}_{\a_{\s(i)}}
     \,R^{\s}_{1\ldots N}(l),\label{F-matrix}
\eea where $P^i_{\a}$ is the embedding of the project operator
$P_{\a}$ in the $i^{{\rm th}}$ space with matric elements
$(P_{\a})_{kl}=\d_{kl}\d_{k\a}$. The sum $\sum^*$ in
(\ref{F-matrix}) is over all non-decreasing sequences of the
labels $\a_{\s(i)}$:
\bea
  && \a_{\s(i+1)}\geq \a_{\s(i)}\quad {\rm if}\quad \s(i+1)>\s(i),\no\\
  && \a_{\s(i+1)}> \a_{\s(i)}\quad {\rm if}\quad
  \s(i+1)<\s(i).\label{Condition}
\eea From (\ref{Condition}), $F_{1\ldots N}(l)$ obviously is a
lower-triangular matrix. Moreover, the F-matrix is non-degenerate
because  all its diagonal elements are non-zero. Using the method
of \cite{Yan06-1},  we find that the F-matrix also satisfies the
factorizing property (\ref{Factorizing}). Hence, the F-matrix
$F_{1\ldots N}(l)$ given by (\ref{F-matrix}) is the desirable
Drinfeld twist.

\subsection{Completely symmetric  representations}
Having found the  F-matrix (\ref{F-matrix}), let us compute the
twisted operators $\tilde{T}_F(l|u)^j_i$ defined by
(\ref{Twisted-Mon-F}). Using the method similar to that in
\cite{Yan06-1}, after a tedious calculation, we obtain the
expressions of $\tilde{T}_F(l|u)^j_i$.  Here we present the
results which are relevant,
\bea
 &&\tilde{T}_F(l|u)^2_2=\frac{\sin(l_{21}-\eta)}{\sin\lt(l_{21}-\eta+
     \eta\langle H,\e_1\rangle\rt)}\otimes_{i}
     \lt(\begin{array}{ll}\frac{\sin(u-z_i)}{\sin(u-z_i+\eta)}&\\
     &1\end{array}\rt)_{(i)},\\[6pt]
 &&\tilde{T}_F(l|u)^2_1=\sum_{i=1}^N\frac{\sin\eta
     \sin(u\hspace{-0.08truecm}-\hspace{-0.08truecm}z_i
     \hspace{-0.08truecm}+\hspace{-0.08truecm}l_{12})}
     {\sin(u\hspace{-0.08truecm}-\hspace{-0.08truecm}z_i
     \hspace{-0.08truecm}+\hspace{-0.08truecm}\eta)\sin l_{12}} E_{12}^i\otimes_{j\neq i}
     \lt(\begin{array}{ll}\frac{\sin(u-z_j)\sin(z_i-z_j+\eta)}{\sin(u-z_j+\eta)\sin(z_i-z_j)}&\\
     &1\end{array}\rt)_{(j)},\\[6pt]
 &&\tilde{T}_F(l|u)^1_2=\frac{\sin(l_{21}\hspace{-0.08truecm}-\hspace{-0.08truecm}\eta)}
     {\sin(l_{21}\hspace{-0.08truecm}+\hspace{-0.08truecm}
     \eta\langle H,\e_1\hspace{-0.08truecm}-\hspace{-0.08truecm}\e_2\rangle)}
     \hspace{-0.08truecm}
     \sum_{i=1}^N\frac{\sin\eta\sin(u\hspace{-0.08truecm}-\hspace{-0.08truecm}z_i
     \hspace{-0.08truecm}+\hspace{-0.08truecm}l_{21}\hspace{-0.08truecm}
     +\hspace{-0.08truecm}\eta\hspace{-0.08truecm}
     +\hspace{-0.08truecm}\eta\langle H,\e_1
     \hspace{-0.08truecm}-\hspace{-0.08truecm}\e_2\rangle)}
     {\sin(u\hspace{-0.08truecm}-\hspace{-0.08truecm}z_i\hspace{-0.08truecm}+\hspace{-0.08truecm}\eta)
     \sin(l_{21}\hspace{-0.08truecm}+\hspace{-0.08truecm}\eta
     \hspace{-0.08truecm}+\hspace{-0.08truecm}\eta\langle
     H,\e_1\hspace{-0.08truecm}-\hspace{-0.08truecm}\e_2\rangle)}\no\\
 &&\quad\quad\quad\quad\quad\quad
     \times E_{21}^i\otimes_{j\neq i} \lt( \begin{array}{ll}
     \frac{\sin(u-z_j)}{\sin(u-z_j+\eta)}&\\
     &\frac{\sin(z_j-z_i+\eta)}{\sin(z_j-z_i)}\end{array}
     \rt)_{(j)},
\eea where $H=\sum_{k=1}^N h^{(k)}$. Applying  the above operators
to the arbitrary  state $|i_1,\ldots,i_N\rangle$ given by
(\ref{Vector-V}), we have
\bea
 &&\tilde{T}_F(m,l|u)^2_2=\frac{\sin(l_{21}-\eta)}{\sin\lt(l_{2}-m_1-\eta\rt)}
     \otimes_{i}
     \lt(\begin{array}{ll}\frac{\sin(u-z_i)}{\sin(u-z_i+\eta)}&\\
     &1\end{array}\rt)_{(i)},\\[6pt]
 &&\tilde{T}_F(m,l|u)^2_1=\sum_{i=1}^N
     \frac{\sin\eta
     \sin(u-z_i+l_{12})}{\sin(u-z_i+\eta)\sin l_{12}}\no\\
 &&\quad\quad\quad\quad\quad\quad
     \times    E_{12}^i \otimes_{j\neq i}
     \lt(\begin{array}{ll}\frac{\sin(u-z_j)\sin(z_i-z_j+\eta)}{\sin(u-z_j+\eta)\sin(z_i-z_j)}&\\
     &1\end{array}\rt)_{(j)},\\[6pt]
 &&\tilde{T}_F(m,l|u)^1_2=\frac{\sin(l_{21}-\eta)}
     {\sin(m_{21}-2\eta)}
     \sum_{i=1}^N\frac{\sin\eta\sin(u-z_i+m_{21}-\eta)}
     {\sin(u-z_i+\eta)\sin(m_{21}-\eta)}\no\\
 &&\quad\quad\quad\quad\quad\quad
     \times E_{21}^i\otimes_{j\neq i} \lt( \begin{array}{ll}
     \frac{\sin(u-z_j)}{\sin(u-z_j+\eta)}&\\
     &\frac{\sin(z_j-z_i+\eta)}{\sin(z_j-z_i)}\end{array}
     \rt)_{(j)}.
\eea It then follows that the two pseudo-particle creation
operators (\ref{Expression-3}) and (\ref{Expression-4}) in the
F-basis simultaneously have the following completely symmetric
polarization free forms:
\bea
 &&\tilde{\T}^-_F(m,\l|u)^2_1=\frac{\sin m_{12}}{\sin(m_1-\l_2)}
   \prod_{k=1}^N\frac{\sin(u+z_k)}{\sin(u+z_k+\eta)}\no\\
  &&\quad\quad\times \sum_{i=1}^N\frac{\sin(\l_1+\xi-z_i)\sin(\l_2+\xi+z_i)\sin2u \sin\eta}
   {\sin(\l_1+\xi+u)\sin(\l_2+\xi+u)\sin(u-z_i+\eta)\sin(u+z_i)}\no\\[6pt]
  &&\quad\quad\quad\quad\quad\quad \times
   E_{12}^i\otimes_{j\neq i}\lt(\begin{array}{ll}
   \frac{\sin(u-z_j)\sin(u+z_j+\eta)\sin(z_i-z_j+\eta)}
   {\sin(u-z_j+\eta)\sin(u+z_j)\sin(z_i-z_j)}&\\
   &1\end{array}\rt)_{(j)},\label{Creation-operator-1}\\[6pt]
 &&\tilde{\T}^+_F(\l,m|u)^1_2=\frac{\sin (m_{21}+\eta)}{\sin(m_2-\l_1)}
   \prod_{k=1}^N\frac{\sin(u+z_k)}{\sin(u+z_k+\eta)}\no\\
  &&\quad\quad\times \sum_{i=1}^N
   \hspace{-0.08truecm}
   \frac{\sin(\l_2\hspace{-0.08truecm}+\hspace{-0.08truecm}\bar{\xi}
   \hspace{-0.08truecm}-\hspace{-0.08truecm}z_i)
   \sin(\l_1\hspace{-0.08truecm}+\hspace{-0.08truecm}\bar{\xi}
   \hspace{-0.08truecm}+\hspace{-0.08truecm}z_i)
   \sin(2u\hspace{-0.08truecm}+\hspace{-0.08truecm}2\eta) \sin\eta}
   {\sin(\l_1\hspace{-0.08truecm}+\hspace{-0.08truecm}\bar{\xi}
   \hspace{-0.08truecm}-\hspace{-0.08truecm}u
   \hspace{-0.08truecm}-\hspace{-0.08truecm}\eta)
   \sin(\l_2\hspace{-0.08truecm}+\hspace{-0.08truecm}\bar{\xi}
   \hspace{-0.08truecm}-\hspace{-0.08truecm}u
   \hspace{-0.08truecm}-\hspace{-0.08truecm}\eta)
   \sin(u\hspace{-0.08truecm}+\hspace{-0.08truecm}z_i)
   \sin(u\hspace{-0.08truecm}-\hspace{-0.08truecm}z_i
   \hspace{-0.08truecm}+\hspace{-0.08truecm}\eta)}\no\\[6pt]
  &&\quad\quad\quad\quad\quad\quad \times
   E_{21}^i\otimes_{j\neq i}\lt(\begin{array}{ll}
   \frac{\sin(u-z_j)\sin(u+z_j+\eta)}
   {\sin(u-z_j+\eta)\sin(u+z_j)}&\\
   &\frac{\sin(z_j-z_i+\eta)}{\sin(z_j-z_i)}\end{array}\rt)_{(j)}.\label{Creation-operator-2}
\eea

Now let us evaluate  the two sets of Bethe states
(\ref{Bethe-state-3}) and (\ref{Bethe-state-4}) in the F-basis (or
the twisted Bethe states)
\bea
 \overline{|v^{(1)}_1,\cdots,v^{(1)}_M\rangle}^{(I)}_F&=&F_{1\ldots N}(\l)
          \,|v^{(1)}_1,\cdots,v^{(1)}_M\rangle^{(I)}_F,\label{Twisted-Bethe-state-1}\\
 \overline{|v^{(2)}_1,\cdots,v^{(2)}_M\rangle}^{(II)}_F&=&F_{1\ldots N}(\l)
          \,|v^{(2)}_1,\cdots,v^{(2)}_M\rangle^{(II)}_F.\label{Twisted-Bethe-state-2}
\eea Since $|1,\ldots,1\rangle$ and $|2,\ldots,2\rangle$ are
invariant under the action of the F-matrix $F_{1\ldots N}(l)$
(\ref{F-matrix}), namely, \bea
 F_{1\ldots N}(l)\,|i,\ldots,i\rangle=|i,\ldots,i\rangle,\quad
 i=1,2,
\eea we have \bea
 &&\overline{|v^{(1)}_1,\cdots,v^{(1)}_M\rangle}^{(I)}_F=
     \tilde{\T}^+_F(\l,\l\hspace{-0.08truecm}+\hspace{-0.08truecm}2\eta\hat{1}
     |v^{(1)}_1)^1_2\cdots
     \tilde{\T}^+_F(\l,\l\hspace{-0.08truecm}+\hspace{-0.08truecm}2M\eta\hat{1}
     |v^{(1)}_M)^1_2|1,\ldots,1\rangle,
   \label{Twisted-Bethe-state-3}\\
 &&\overline{|v^{(2)}_1,\cdots,v^{(2)}_M\rangle}^{(II)}_F =
   \tilde{\T}^-_F(\l\hspace{-0.08truecm}-\hspace{-0.08truecm}2\eta\hat{2},\l
   |v^{(2)}_1)^2_1
   \cdots
   \tilde{\T}^-_F(\l\hspace{-0.08truecm}-\hspace{-0.08truecm}2M\eta\hat{2},\l
   |v^{(2)}_M)^2_1
   |2,\ldots,2\rangle.
   \label{Twisted-Bethe-state-4}
\eea Thanks to the polarization free representations
(\ref{Creation-operator-1}) and (\ref{Creation-operator-2}) of the
pseudo-particle creation operators,  we also obtain completely
symmetric expressions of the two sets of Bethe sates in the
F-basis: \bea
 \overline{|v^{(1)}_1,\cdots,v^{(1)}_M\rangle}^{(I)}_F&=&\prod_{k=1}^M
     \lt\{ \frac{\sin(\l_{12}-\eta+2k\eta)}{\sin(\l_{12}+k\eta)}
     \prod_{n=1}^{N}\frac{\sin(v^{(1)}_k-z_n)}{\sin(v^{(1)}_k-z_n+\eta)}\rt\}\no\\
 &&\quad\quad \times \sum_{i_1<i_2\ldots<i_M}B^{(I)}_M\lt(\{v^{(1)}_{\a}\}|\{z_{i_n}\}\rt)
     E^{i_1}_{21}\ldots
     E^{i_M}_{21}\,|1,\ldots,1\rangle,\label{BA-state-1}\\
 \overline{|v^{(2)}_1,\cdots,v^{(2)}_M\rangle}^{(II)}_F&=&\prod_{k=1}^M
     \lt\{ \frac{\sin(\l_{12}+2k\eta)}{\sin(\l_{12}+k\eta)}
     \prod_{n=1}^{N}\frac{\sin(v^{(2)}_k+z_n)}{\sin(v^{(2)}_k+z_n+\eta)}\rt\}\no\\
 &&\quad\quad \times \sum_{i_1<i_2\ldots<i_M}B^{(II)}_M\lt(\{v^{(2)}_{\a}\}|\{z_{i_n}\}\rt)
     E^{i_1}_{12}\ldots
     E^{i_M}_{12}\,|2,\ldots,2\rangle.\label{BA-state-2}
\eea Here the functions $B^{(I)}_M\lt(\{v_{\a}\}|\{z_{i_n}\}\rt)$
and $B^{(II)}_M\lt(\{v_{\a}\}|\{z_{i_n}\}\rt)$ are given by \bea
 &&B^{(I)}_M\lt(\{v_{\a}\}|\{z_{i_n}\}\rt)=\prod_{n=1}^M\prod_{k=1}^M
   \frac{\sin(v_n-z_{i_k}+\eta)\sin(v_n+z_{i_k})}{\sin(v_n-z_{i_k})\sin(v_n+z_{i_k}+\eta)}\no\\
 &&\,\quad\times \sum_{\s\in\mathcal{S}_M}\prod_{n=1}^M\hspace{-0.08truecm}
   \lt\{\frac{\sin(\l_2\hspace{-0.08truecm}+\hspace{-0.08truecm}\bar{\xi}
   \hspace{-0.08truecm}-\hspace{-0.08truecm}z_{i_{\s(n)}})
   \sin(\l_1\hspace{-0.08truecm}+\hspace{-0.08truecm}\bar{\xi}
   \hspace{-0.08truecm}+\hspace{-0.08truecm}z_{i_{\s(n)}})
   \sin(2v_n\hspace{-0.08truecm}+\hspace{-0.08truecm}2\eta)\sin\eta}
   {\sin(\l_2\hspace{-0.08truecm}+\hspace{-0.08truecm}\bar{\xi}
   \hspace{-0.08truecm}-\hspace{-0.08truecm}v_n
   \hspace{-0.08truecm}-\hspace{-0.08truecm}\eta)
   \sin(\l_1\hspace{-0.08truecm}+\hspace{-0.08truecm}\bar{\xi}
   \hspace{-0.08truecm}-\hspace{-0.08truecm}v_n
   \hspace{-0.08truecm}-\hspace{-0.08truecm}\eta)
   \sin(v_n\hspace{-0.08truecm}+\hspace{-0.08truecm}z_{i_{\s(n)}})
   \sin(v_n\hspace{-0.08truecm}-\hspace{-0.08truecm}z_{i_{\s(n)}}
   \hspace{-0.08truecm}+\hspace{-0.08truecm}\eta)}
   \rt.\no\\
 &&\,\quad\quad \times \prod_{k>n}^M\lt.
   \frac{\sin(v_k\hspace{-0.08truecm}-\hspace{-0.08truecm}z_{i_{\s(n)}})
    \sin(v_k\hspace{-0.08truecm}+\hspace{-0.08truecm}z_{i_{\s(n)}}
    \hspace{-0.08truecm}+\hspace{-0.08truecm}\eta)
    \sin(z_{i_{\s(k)}}\hspace{-0.08truecm}-\hspace{-0.08truecm}z_{i_{\s(n)}}
    \hspace{-0.08truecm}+\hspace{-0.08truecm}\eta)}
   {\sin(v_k\hspace{-0.08truecm}-\hspace{-0.08truecm}z_{i_{\s(n)}}
   \hspace{-0.08truecm}+\hspace{-0.08truecm}\eta)
   \sin(v_k\hspace{-0.08truecm}+\hspace{-0.08truecm}z_{i_{\s(n)}})
   \sin(z_{i_{\s(k)}}\hspace{-0.08truecm}-\hspace{-0.08truecm}z_{i_{\s(n)}})}
   \rt\},\label{Function-B-1}\\[6pt]
 &&B^{(II)}_M\lt(\{v_{\a}\}|\{z_{i_n}\}\rt)=\hspace{-0.18truecm}
   \sum_{\s\in\mathcal{S}_M}\prod_{n=1}^M\hspace{-0.08truecm}
   \lt\{\frac{\sin(\l_1\hspace{-0.08truecm}+\hspace{-0.08truecm}\xi
   \hspace{-0.08truecm}-\hspace{-0.08truecm}z_{i_{\s(n)}})
   \sin(\l_2\hspace{-0.08truecm}+\hspace{-0.08truecm}\xi
   \hspace{-0.08truecm}+\hspace{-0.08truecm}z_{i_{\s(n)}})
   \sin(2v_n)\sin\eta}
   {\sin(\l_1\hspace{-0.08truecm}+\hspace{-0.08truecm}\xi
   \hspace{-0.08truecm}+\hspace{-0.08truecm}v_n)
   \sin(\l_2\hspace{-0.08truecm}+\hspace{-0.08truecm}\xi
   \hspace{-0.08truecm}+\hspace{-0.08truecm}v_n)
   \sin(v_n\hspace{-0.08truecm}-\hspace{-0.08truecm}z_{i_{\s(n)}}
   \hspace{-0.08truecm}+\hspace{-0.08truecm}\eta)
   \sin(v_n\hspace{-0.08truecm}+\hspace{-0.08truecm}z_{i_{\s(n)}})}
   \rt.\no\\
   &&\,\quad\quad \times \prod_{k>n}^M\lt.
   \frac{\sin(v_n-z_{i_{\s(k)}}) \sin(v_n+z_{i_{\s(k)}}+\eta) \sin(z_{i_{\s(n)}}-z_{i_{\s(k)}}+\eta)}
   {\sin(v_n-z_{i_{\s(k)}}+\eta) \sin(v_n+z_{i_{\s(k)}}) \sin(z_{i_{\s(n)}}-z_{i_{\s(k)}})}
   \rt\}.\label{Function-B-2}
\eea We remark that if the parameters $\{v^{(1)}_k\}$ (or
$\{v^{(2)}_k\}$) do not satisfy the associated Bethe ansatz
equations (\ref{BA-D-1})(or (\ref{BA-D-2})), the corresponding
twisted states (\ref{Twisted-Bethe-state-3}) and
(\ref{Twisted-Bethe-state-4}) become off-shell Bethe states. These
off-shell Bethe states can still be expressed in the same forms as
those of (\ref{BA-state-1})-(\ref{Function-B-2}) (but the
corresponding parameters are not necessarily the roots of the
Bethe ansatz equations).


\section{ Conclusions}
\label{C} \setcounter{equation}{0}

We have constructed the factorizing F-matrix (\ref{F-matrix}) in
the face picture for the open XXZ chain with non-diagonal boundary
terms, where the non-diagonal K-matrices $K^{\pm}(u)$ are given by
(\ref{K-matrix-2-1}) and (\ref{K-matrix-6}). It is found that in
the F-basis the pseudo-particle creation operators, which generate
the complete set of the eigenstates of the model, simultaneously
take the completely symmetric and polarization free forms
(\ref{Creation-operator-1}) and (\ref{Creation-operator-2}). This
allows us to obtain the explicit and completely symmetric
expressions (\ref{BA-state-1}) and (\ref{BA-state-2}) of the two
sets of (off-shell) Bethe states.

The results of this paper make it feasible to derive the
determinant representations of partition functions and scalar
products of the complete Bethe states for the open XXZ chain with
non-diagonal boundary terms specified by the non-diagonal
K-matrices $K^{\pm}(u)$ (\ref{K-matrix-2-1}) and
(\ref{K-matrix-6}). These results will be presented elsewhere
\cite{Yan09}.

\section*{Acknowledgements}
The financial support from  Australian Research Council is
gratefully acknowledged.


\section*{Appendix A: Bethe states (\ref{Bethe-state-1}) }
\setcounter{equation}{0}
\renewcommand{\theequation}{A.\arabic{equation}}

In this appendix, we will prove that if the parameters
$\{v^{(1)}_k\}$ satisfy the first set of Bethe ansatz equations
(\ref{BA-D-1}) the Bethe states (\ref{Bethe-state-1}) give rise to
the eigenvalues (\ref{Eigenfuction-D-1}) of the transfer matrix.

Using the QYBE (\ref{QYB}), the dual RE (\ref{DRE-V}) and the
definition of the monodromy matrix $\mathbb{T}^+(u)$
(\ref{Mon-V-1}), we can show  that $\mathbb{T}^+(u)$ satisfies the
following relation \bea
 &&\R_{2,1}(u_2-u_1)\lt(\mathbb{T}_1^+(u_1)\rt)^{t_1}
   \R_{1,2}(-u_1-u_2-2\eta)\lt(\mathbb{T}_2^+(u_2)\rt)^{t_2}\no\\
 &&\qquad\qquad =\lt(\mathbb{T}_2^+(u_2)\rt)^{t_2}
   \R_{2,1}(-u_1-u_2-2\eta)\lt(\mathbb{T}_1^+(u_1)\rt)^{t_1}
   \R_{1,2}(u_2-u_1).\no
\eea This, together with (\ref{Face-vertex}) and
(\ref{Face-vertex3})-(\ref{Face-vertex4}), implies that the
operators $\T^+(m|u)$ defined by (\ref{Mon-F-1}) satisfy the
relations \bea
 &&\sum_{i_1,j_1}\sum_{i_2,j_2}R(u_2-u_1;m)^{i_0\,j_0}_{i_1\,j_1}
       R(-u_1-u_2-2\eta;m)^{j_1\,i_2}_{j_2\,i_3}\no\\
 &&\qquad\qquad\times
       \T^+(m+\eta(\hat{\imath}_3+\hat{\jmath}_2)|u_2)^{j_2}_{j_3}
       \T^+(m+\eta(\hat{\imath}_1+\hat{\jmath}_1)|u_1)^{i_1}_{i_2}\no\\
 &&\quad\quad=
   \sum_{i_1,j_1}\sum_{i_2,j_2}R(-u_1-u_2-2\eta;m)^{i_0\,j_1}_{i_1\,j_2}
       R(u_2-u_1;m)^{j_2\,i_2}_{j_3\,i_3}\no\\
 &&\qquad\qquad\times
       \T^+(m+\eta(\hat{\imath}_1+\hat{\jmath}_2)|u_1)^{i_1}_{i_2}
       \T^+(m+\eta(\hat{\imath}_0+\hat{\jmath}_0)|u_2)^{j_0}_{j_1}.
       \label{Exchange-relation-1}
\eea Following \cite{Yan07}, let us introduce operators
$\A^{+(1)}$ and $\D^{+(1)}$
\bea
 \D^{+(1)}(m|u)&=&\T^+(m|u)^2_2,\no\\
 \A^{+(1)}(m|u)&=&\frac{\sin(m_{12}-\eta)}{\sin m_{12}}
       \lt\{\T^+(m|u)^1_1-R(-2u-2\eta;m+\eta\hat{2})^{1\,2}_{2\,1}
       \T^+(m|u)^2_2\rt\}.\no
\eea We may check that the transfer matrix $\tau(u)$ (\ref{trans})
can be expressed in terms of a linear combination of the above
operators \bea
 \tau(u)&=&tr\lt(K^+(u)\mathbb{T}(u)\rt)
          =tr\lt(\lt(K^-(u)\rt)^{t_0}\lt(\mathbb{T}^+(u)\rt)^{t_0}\rt)
          =tr\lt(K^-(u)\mathbb{T}^+(u)\rt)\no\\
 &=&\frac{\sin(\l_1+\xi-u)}{\sin(\l_1+\xi+u)}\A^{+(1)}(\l|u)\no\\
 &&\qquad +\frac{\sin(\l_1+\xi+u+\eta)\,\sin(\l_2+\xi-u-\eta)\,\sin(2u)}
          {\sin(\l_1+\xi+u)\,\sin(\l_2+\xi+u)\,\sin(2u+\eta)}
          \D^{+(1)}(\l|u).
\eea Using the technique developed in \cite{Yan04}, after tedous
calculations, we find that the state $|\O^{(I)}(\l)\rangle$  given
by (\ref{Vac-1}) is a reference state  in the usual  sense,
\bea
 \A^{+(1)}(\l+N\eta\hat{1}|u)\,|\O^{(I)}(\l)\rangle&=&
     \frac{\sin(\l_2+\bar{\xi}-u)\,\sin(\l_1+\bar{\xi}+u)}
     {\sin(\l_2+\bar{\xi}-u-\eta)\,\sin(\l_1+\bar{\xi}-u-\eta)}\no\\
 &&\quad\quad \times\frac{\sin(2u+2\eta)}{\sin(2u+\eta)}\,|\O^{(I)}(\l)\rangle,\\
 \D^{+(1)}(\l+N\eta\hat{1}|u)\,|\O^{(I)}(\l)\rangle&=&
     \prod_{l=1}^{N}\frac{\sin(u+z_l)\,\sin(u-z_l)}
     {\sin(u+z_l+\eta)\sin(u-z_l+\eta)}\no\\
 &&\quad\quad \times \frac{\sin(\l_2+\bar{\xi}+u+\eta)}
     {\sin(\l_2+\bar{\xi}-u-\eta)}\,|\O^{(I)}(\l)\rangle,\\
 \T^+(\l+N\eta\hat{1}|u)^1_2\,|\O^{(I)}(\l)\rangle&\neq&0,\\
 \T^+(\l+N\eta\hat{1}|u)^2_1\,|\O^{(I)}(\l)\rangle&=&0.
\eea Moreover, by (\ref{Exchange-relation-1}) we obtain  the
commutation relations,
\bea
 &&\A^{+(1)}(m+\eta(\hat{1}+\hat{2})|u_1)\T^+(m+2\eta\hat{1}|u_2)^1_2\no\\
 &&\quad=
    \frac{\sin(u_2-u_1+\eta)\,\sin(u_2+u_1)}{\sin(u_2-u_1)\,\sin(u_2+u_1+\eta)}
    \T^+(m+2\eta\hat{1}|u_2)^1_2\A^{+(1)}(m+2\eta\hat{1}|u_1)\no\\
 &&\quad\quad-
    \frac{\sin(\eta)\sin(2u_1)\sin(u_2-u_1+\l_{21}-\eta)}
    {\sin(u_2-u_1)\sin(2u_1+\eta)\sin(\l_{21}-\eta)}
    \T^+(m\hspace{-0.1truecm}+\hspace{-0.1truecm}2\eta\hat{1}|u_1)^1_2
    \A^{+(1)}(m\hspace{-0.1truecm}+\hspace{-0.1truecm}2\eta\hat{1}|u_2)\no\\
 &&\quad\quad-
    \frac{\sin(\eta)\sin(2u_2+2\eta)\sin(2u_1)\sin(u_2-u_1+\l_{21})}
    {\sin(u_1+u_2+\eta)\sin(2u_2+\eta)\sin(2u_1+\eta)\sin(\l_{21}-\eta)}\no\\
 &&\qquad\qquad\qquad\qquad\times   \T^+(m+2\eta\hat{1}|u_1)^1_2\D^{+(1)}(m+2\eta\hat{1}|u_2),
    \label{Re-relation-1}\\
 &&\D^{+(1)}(m+2\eta\hat{2}|u_2)\T^+(m+\eta(\hat{1}+\hat{2})|u_1)^1_2\no\\
 &&\quad=
    \frac{\sin(u_2\hspace{-0.1truecm}+\hspace{-0.1truecm}u_1\hspace{-0.1truecm}+\hspace{-0.1truecm}2\eta)
    \sin(u_2\hspace{-0.1truecm}-\hspace{-0.1truecm}u_1\hspace{-0.1truecm}+\hspace{-0.1truecm}\eta)}{\sin(u_2+u_1+\eta)\sin(u_2-u_1)}
    \T^+(m\hspace{-0.1truecm}+\hspace{-0.1truecm}\eta(\hat{1}\hspace{-0.1truecm}+\hspace{-0.1truecm}\hat{2})|u_1)^1_2
    \D^{+(1)}(m\hspace{-0.1truecm}+\hspace{-0.1truecm}\eta(\hat{1}\hspace{-0.1truecm}+\hspace{-0.1truecm}\hat{2})|u_2)\no\\
 &&\quad\quad-
    \frac{\sin(\eta)\,\sin(2u_1+2\eta)\,\sin(u_2-u_1+\l_{21}+\eta)}
    {\sin(u_2-u_1)\,\sin(2u_1+\eta)\,\sin(\l_{21}+\eta)}\no\\
 &&\qquad\qquad\qquad\qquad\times
    \T^+(m+\eta(\hat{1}+\hat{2})|u_2)^1_2\D^{+(1)}(m+\eta(\hat{1}+\hat{2})|u_1)\no\\
 &&\quad\quad-
    \frac{\sin(\eta)\sin(u_2+u_1+\l_{21}+2\eta)}
    {\sin(u_1+u_2+\eta)\sin(\l_{12}-\eta)}\no\\
 &&\qquad\qquad\qquad\qquad\times
    \T^+(m+\eta(\hat{1}+\hat{2})|u_2)^1_2\A^{+(1)}(m+\eta(\hat{1}+\hat{2})|u_1).
    \label{Re-relation-2}
\eea Carrying out the generalized Bethe ansatz \cite{Yan04}, we
finally find that the Bethe states
$|v^{(1)}_1,\cdots,v^{(1)}_M\rangle^{(I)}$ (\ref{Bethe-state-1})
give rise to the first set of eigenvalues $\L^{(1)}(u)$
(\ref{Eigenfuction-D-1}) provided that the parameters
$\{v^{(1)}_k\}$ satisfy the first set of Bethe ansatz equations
(\ref{BA-D-1}).



\begin{thebibliography}{99}
\bibitem{Kor93} V.\,E. Korepin, N.\,M. Bogoliubov and A.\,G.
Izergin, {\it Quantum Inverse Scattering Method and Correlation
Functions}, Cambridge University Press, 1993.
\bibitem{Mai00} J.\,M. Maillet and J. Sanchez de Santos, Drinfeld
    twists and algebraic Bethe ansatz, {\it Amer. Math. Soc.
    Transl.\/} {\bf 201} (2000), 137.
\bibitem{Dri83} V.\,G. Drinfeld, {\it Sov. Math. Dokl.\/} {\bf 28}
(1983), 667.
\bibitem{Kit99} N. Kitanine, J.\,M. Maillet and V. Terras, {\it
     Nucl. Phys.\/} {\bf B 554} (1999), 647.
\bibitem{Ter99} V. Terras, {\it Lett. Math. Phys.\/} {\bf 48}
(1999), 263.
\bibitem{Alb00} T.\,-D. Albert, H. Boos, R. Flume and K. Rulig,
{\it J. Phys.\/} {\bf A 33} (2000), 4963.
\bibitem{Alb00-1}  T.\,-D. Albert, H. Boos, R. Flume,
R.\,H. Poghossian and K. Rulig, {\it Lett. Math. Phys.\/} {\bf 53}
(2000), 201.
\bibitem{Alb01} T.\,-D. Albert and K. Rulig, {\it J. Phys.\/} {\bf
A 34} (2001), 1569.
\bibitem{Zha06} S.\,-Y. Zhao, W.\,-L. Yang and Y.\,-Z. Zhang, {\it
J. stat. Mech.\/} (2005), {\bf P04005}; {\it Commun. Math. Phys.}
{\bf 268} (2006), 505; {\it Int. J. Mod.  Phys.\/} {\bf B 20}
(2006), 505.
\bibitem{Yan06-1} W.\,-L. Yang, Y.\,-Z. Zhang and S.\,-Y. Zhao,
{\it JHEP\/} {\bf 12} (2004), 038; {\it Commun. Math. Phys.} {\bf
264} (2006), 87.
\bibitem{Wan02} Y.\,-S. Wang, {\it Nucl. Phys.\/} {\bf B 622}
(2002), 633.
\bibitem{Kit07} N. Kitanine, K.\,K. Kozlowski, J.\,M. Maillet, G.
     Niccoli, N.\, A. Slavnov and V. Terras, {\it J. Stat. Mech.\/}
     (2007), {\bf P10009}.
\bibitem{Skl88} E.\,K. Sklyanin, {\it J. Phys. \/} {\bf A 21}
(1988), 2375.
\bibitem{Nep04} R.\,I. Nepomechie,
{\it J. Stat. Phys.\/} {\bf 111} (2003), 1363; {\it J. Phys.\/}
{\bf A 37} (2004), 433.
\bibitem{Cao03} J. Cao, H.\,-Q. Lin, K.\,-J. Shi and Y. Wang, {\it
Nucl. Phys.\/} {\bf B 663} (2003), 487.
\bibitem{Yan04} W.\,-L. Yang and R. Sasaki, {\it Nucl. Phys.\/}
    {\bf B 679} (2004), 495; {\it J. Math. Phys.\/} {\bf 45} (2004),
    4301; W.\,-L. Yang, R. Sasaki and Y.\,-Z. Zhang, {\it JHEP}
    {\bf 09} (2004), 046.
\bibitem{Gal05} W. Galleas and M.\,J. Martins, {\it Phys. Lett.\/}
{\bf A 335} (2005), 167; C.\,S. Melo, G.\,A.\,P. Ribeiro and
M.\,J. Martins, {\it Nucl. Phys.\/} {\bf B 711} (2005), 565.
\bibitem{Gie05}J. de Gier and P. Pyatov, {\it J. Stat. Mech.\/}
(2004), {\bf P03002}; A. Nichols, V. Rittenberg and J. de Gier,
{\it J. Stat. Mech.\/} (2005), {\bf P05003}; J. de Gier, A.
Nichols, P. Pyatov and V. Rittenberg, {\it Nucl. Phys.\/} {\bf B
729} (2005), 387.
\bibitem{Gie05-1} J. de Gier and F.\,H.\,L. Essler, {\it Phys.
Rev. Lett.\/} {\bf 95} (2005), 240601; {\it J. Stat. Mech.\/}
(2006), {\bf P 12011}.
\bibitem{Yan04-1} W.\,-L. Yang, Y.\,-Z. Zhang and M. Gould, {\it
Nucl. Phys.\/} {\bf B 698} (2004), 312.
\bibitem{Baj06} Z. Bajnok, {\it J. Stat. Mech.\/} (2006), {\bf P06010}.
\bibitem{Yan05}  W.\,-L. Yang and  Y.\,-Z. Zhang, {\it JHEP} {\bf
01} (2005), 021; W.\,-L. Yang, Y.\,-Z. Zhang and R. Sasaki, {\it
Nucl. Phys.\/} {\bf 729} (2005), 594.
\bibitem{Doi06} A. Doikou and P.\,P. Martin, {\it J. Stat.
Mech.\/} (2006), {\bf P06004}; A. Dikou, {\it J. Stat. Mech.\/}
(2006), {\bf P09010}.
\bibitem{Mur06} R. Murgan, R.\,I. Nepomechie and C. Shi, {\it J.
Stat. Mech.\/}   (2006) {\bf P08006}.
\bibitem{Bas07} P. Baseilhac and K. Koizumi, {\it J. Stat.
Mech.\/} (2007), {\bf P09006}.
\bibitem{Gal08} W. Galleas, {\it Nucl. Phys.\/} {\bf B 790}
(2008), 524.
\bibitem{Mur09} R. Murgan, {\it JHEP} {\bf 04} (2009), 076.
\bibitem{Yan07} W.\,-L. Yang and Y.\,-Z. Zhang, {\it JHEP} {\bf 04}
 (2007), 044; {\it Nucl. Phys.\/} {\bf B 789} (2008), 591.
\bibitem{Nep03} R.\,I. Nepomechie and F. Ravanini, {\it J.
Phys.\/} {\bf A 36} (2003), 11391; Addendum, {\it J. Phys. \/}
{\bf A 37} (2004), 1945.
\bibitem{Yan06} W.\,-L. Yang, R.\,I. Nepomechie and Y.\,-Z. Zhang,
{\it Phys. Lett.\/} {\bf B 633} (2006), 664; W.\,-L. Yang, and
Y.\,-Z. Zhang, {\it Nucl. Phys.\/} {\bf B 744} (2006), 312; L.
Frappat, R.\,I. Nepomechie and E. Ragoucy, {\it J. Stat. Mech.}
(2007), {\bf P09008}.
\bibitem{Veg93} H.\,J. de Vega and A. Gonzalez-Ruiz, {\it J. Phys.
\/} {\bf A 26} (1993), L519.
\bibitem{Gho94} S. Ghoshal and A.\,B. Zamolodchikov, {\it Int. J.
Mod. Phys.\/} {\bf A 9} (1994), 3841.
\bibitem{Bax82} R.\,J. Baxter, {\it Exactly solved models in
statistical mechanics}, Academic Press, New York, 1982.
\bibitem{Yan09} W.\,-L. Yang, X. Chen, J. Feng, K. Hao, B.-Y. Hou,
K.\,-J. Shi and Y.\,-Z. Zhang, Determinant representations of scalar products for the open
XXZ chain with non-diagonal boundary terms.
\bibitem{Fel96} G. Felder and A. Varchenko, {\it Nucl. Phys.\/}
   {\bf B 480} (1996), 485.
\bibitem{Hou03} B.\,Y. Hou, R. Sasaki and W.\,-L. Yang, {\it Nucl.
   Phys.\/} {\bf B 663} (2003), 467; {\it J. Math. Phys.\/} {\bf
   45} (2004), 559.

\end{thebibliography}
\end{document}